\documentclass[sigconf, nonacm]{acmart}

\newcommand\vldbdoi{10.14778/3421424.3421431}
\newcommand\vldbpages{XXX-XXX}
\newcommand\vldbvolume{14}
\newcommand\vldbissue{1}
\newcommand\vldbyear{2021}
\newcommand\vldbauthors{\authors}
\newcommand\vldbtitle{\shorttitle} 
\newcommand\vldbavailabilityurl{}
\newcommand\vldbpagestyle{empty} 

\usepackage{balance}  % for  \balance command ON LAST PAGE  (only there!)
\usepackage{booktabs}
\usepackage[shortlabels]{enumitem}
\usepackage{multirow}

\setlist[itemize]{leftmargin=*}
\setlist[enumerate]{leftmargin=*}

\newcommand{\yuliang}[1]{{\it\small\textcolor{blue}{[[[ {#1}\ --yuliang ]]]}}}
\newcommand{\wctan}[1]{{\it\small\textcolor{red}{[[[ {#1}\ --wangchiew ]]]}}}
\newcommand{\jinfeng}[1]{{\it\small\textcolor{cyan}{[[[ {#1}\ --jinfeng ]]]}}}
\newcommand{\yoshi}[1]{{\it\small\textcolor{mygreen}{[[[ {#1}\ --yoshi ]]]}}}

\newcommand{\system}{{\sc Ditto}}
\newcommand{\rev}[1]{{{#1}}}

\begin{document}

\title{Deep Entity Matching with Pre-Trained Language Models}

% \authors{Yuliang Li, Jinfeng Li, Yoshihiko Suhara, AnHai Doan, Wang-Chiew Tan}

\author{Yuliang Li, Jinfeng Li, Yoshihiko Suhara}
\affiliation{%
  \institution{Megagon Labs}
%   \streetaddress{444 Castro St Suite 720}
%   \city{Mountain View}
%   \state{California, 94041}
}
\email{{yuliang,jinfeng,yoshi}@megagon.ai}

\author{AnHai Doan}
\affiliation{%
  \institution{University of Wisconsin Madison}
%   \streetaddress{1210 W. Dayton St}
%   \city{Madison}
%   \state{Wisconsin, 53706}
}
\email{anhai@cs.wisc.edu}

\author{Wang-Chiew Tan}
\affiliation{%
  \institution{Megagon Labs}
%   \streetaddress{444 Castro St Suite 720}
%   \city{Mountain View}
%   \state{California, 94041}
}
\email{wangchiew@megagon.ai}

\begin{abstract}
% Entity Matching (EM) refers to the problem of deciding whether two data entries refer to the same real-world entity.
%and more generally, it refers to the problem of %determining matching data entries in different %datasets.

%Given an input of two structured data items, 
%Entity Matching (EM) decides whether the two items
%refer to the same real-world entity. 
%As one of the most fundamental tasks in data %integration,
%EM has a wide range of applications from entity %similarity search to
%joining/merging database tables from different %sources.
%Existing solutions for EM include both rule-based %and learning-based methods
%with the recently proposed deep-learning-based %methods delivering
%the state-of-the-art (SOTA) matching qualities.
%However, the success of these models requires a %significant amount of training data
%and they lack the ability of learning beyond the %training data.

We present \system, a novel entity matching system based on pre-trained Transformer-based language models. 
We fine-tune and cast EM as a sequence-pair classification problem to leverage such models with a simple architecture. Our experiments show that a straightforward application of language models 
such as  
BERT, DistilBERT, or \rev{RoBERTa} pre-trained on large text corpora 
already significantly improves the matching quality
and outperforms previous state-of-the-art (SOTA), by up to \rev{29\%} of F1 score on benchmark datasets. 
We also developed three
optimization techniques to further improve \system's matching capability.
\system{} allows domain knowledge to be injected by highlighting important pieces of input information that may be of interest when making matching decisions. \system{} also summarizes strings that are too long so that only the essential information is retained and used for EM. 
%These two techniques help \system{} to pay attention to the right information when making entity matching decisions.
Finally, \system{} adapts a SOTA technique on data augmentation for text to EM to augment the training data with (difficult) examples. This way, \system{} is forced to learn ``harder'' to improve the model's matching capability. The optimizations we developed 
further boost the performance of \system\
by up to \rev{9.8\%}. Perhaps more surprisingly, 
we establish that \system{} can achieve the previous SOTA results with at most half the number of labeled data. 
Finally, we demonstrate \system's effectiveness on
a real-world large-scale EM task. On matching two company datasets consisting of
789K and 412K records, \system\ achieves a high F1 score of 96.5\%.
\end{abstract}

\maketitle

%%% do not modify the following VLDB block %%
%%% VLDB block start %%%
\pagestyle{\vldbpagestyle}
\begingroup\small\noindent\raggedright\textbf{PVLDB Reference Format:}\\
\vldbauthors. \vldbtitle. PVLDB, \vldbvolume(\vldbissue): \vldbpages, \vldbyear.\\
\href{https://doi.org/\vldbdoi}{doi:\vldbdoi}
\endgroup
\begingroup
\renewcommand\thefootnote{}\footnote{\noindent
This work is licensed under the Creative Commons BY-NC-ND 4.0 International License. Visit \url{https://creativecommons.org/licenses/by-nc-nd/4.0/} to view a copy of this license. For any use beyond those covered by this license, obtain permission by emailing \href{mailto:info@vldb.org}{info@vldb.org}. Copyright is held by the owner/author(s). Publication rights licensed to the VLDB Endowment. \\
\raggedright Proceedings of the VLDB Endowment, Vol. \vldbvolume, No. \vldbissue\ %
ISSN 2150-8097. \\
\href{https://doi.org/\vldbdoi}{doi:\vldbdoi} \\
}\addtocounter{footnote}{-1}\endgroup
%%% VLDB block end %%%

%%% do not modify the following VLDB block %%
%%% VLDB block start %%%
\ifdefempty{\vldbavailabilityurl}{}{
\vspace{.3cm}
\begingroup\small\noindent\raggedright\textbf{PVLDB Artifact Availability:}\\
The source code, data, and/or other artifacts have been made available at \url{\vldbavailabilityurl}.
\endgroup
}

\section{Introduction}\label{sec:intro}

\begin{figure*}[!ht]
    \centering
    \includegraphics[width=0.90\textwidth]{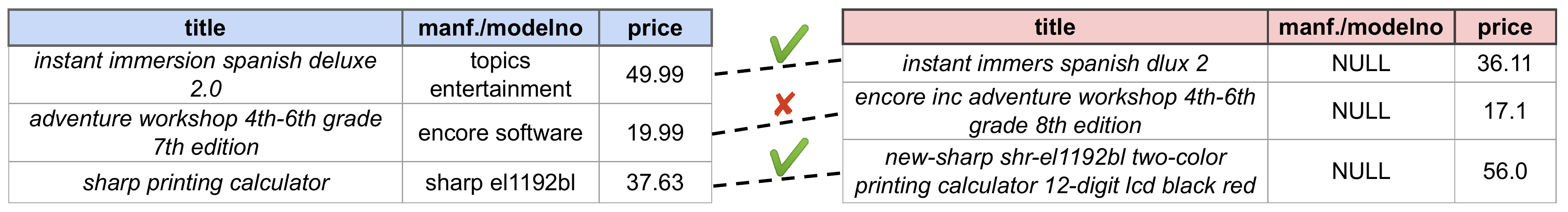}
    \vspace{-4mm}
    \caption{\small Entity Matching: determine the matching entries from two datasets.}
    \label{fig:EMExample}
    \vspace{-4mm}
\end{figure*}

Entity Matching (EM) refers to the problem of determining
whether two data entries refer to the same real-world entity. 
%It is the fundamental building block for finding %matching entries in different datasets,
%which is a pervasive problem in many enterprises %and organizations. 
Consider the two datasets about products in Figure~\ref{fig:EMExample}.
%. The examples are drawn from 
%the Entity-Resolution %benchmark~\cite{kopcke2010evaluation} and the %Magellan data repository~\cite{magellandata}
% \wctan{add reference} 
%but slightly modified to fit the space and %discussion here.
The goal is to determine the set of pairs of data entries, one entry from each table so that each pair of entries refer to the same product. %\yuliang{entity $\rightarrow$ product?}
%Typically, an entry in one table matches at most one entry in the other table. However, in general, an entry in one table can match several entries in the other table, especially if duplicate entries exist. 

%Regardless of whether duplicates exist, 
If the datasets are large, it can be expensive to determine the pairs of matching entries. For this reason, EM is typically accompanied by a pre-processing step, called {\em blocking}, to prune pairs of entries that are unlikely matches to reduce the number of candidate pairs to consider. 
As we will illustrate, correctly {\em matching} the candidate pairs requires substantial language understanding 
%(e.g., ``immersion'' is likely the same as ``immers'' in this context) 
and domain-specific knowledge.
%(e.g., the grade level and edition numbers are important information for books). 
Hence, entity matching remains a challenging task even for the most advanced EM solutions.
%\yuliang{added a few sentences. Pls check.}
% \yuliang{TODO: add the challenges from the examples.}

\begin{figure}[!ht]
    \vspace{-1mm}
\centering
\includegraphics[width=0.47\textwidth]{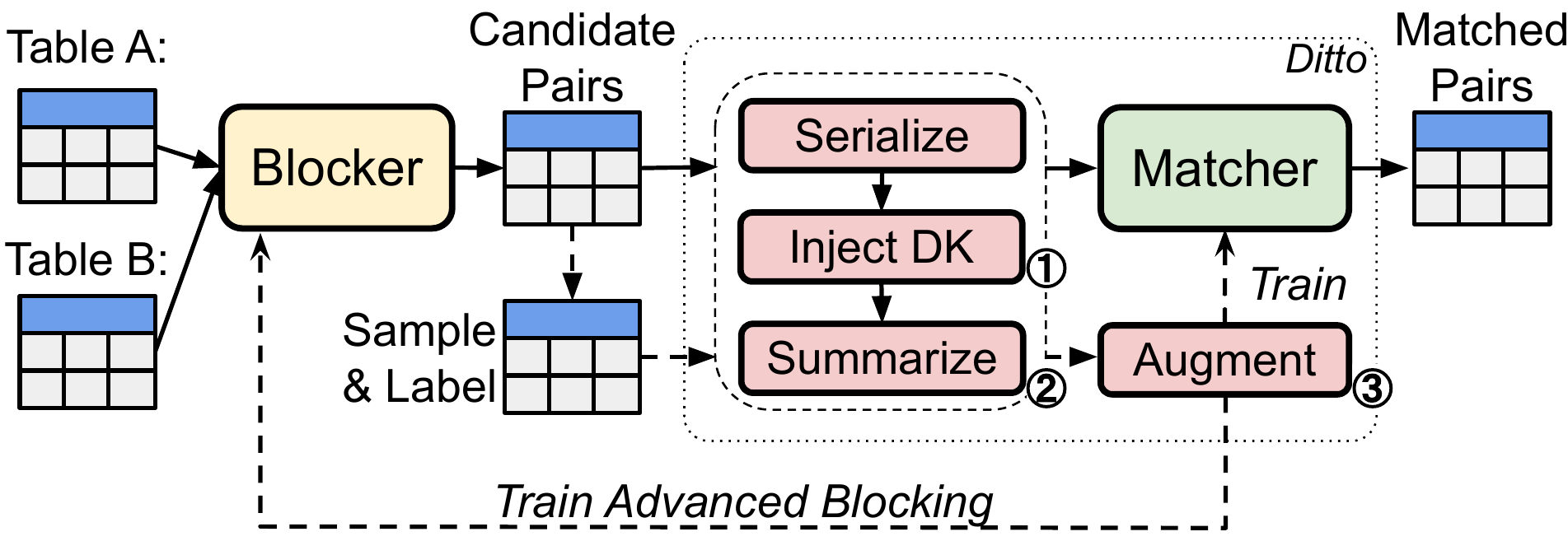}
    \vspace{-2mm}
\caption{\small An EM system architecture with \system{} as the matcher. In addition to the training data,
the user of \system\ can specify (1) a method for injecting domain knowledge (DK),
(2) a summarization module for keeping the essential information, and
(3) a data augmentation (DA) operator to strengthen the training set.}
\label{fig:pipeline}
    \vspace{-2mm}
\end{figure}

We present \system, a novel EM solution based on pre-trained \\ Transformer-based language models (or {\em pre-trained language models} in short). We cast EM as a sequence-pair classification problem to leverage such models, which have been shown to generate highly contextualized embeddings that capture better language understanding compared to traditional word embeddings.
\system{} further improves its matching capability through three optimizations: (1) It allows domain knowledge to be added by highlighting important pieces
of the input that may be useful for matching decisions.
(2) It summarizes long strings so that only the most essential information is retained and used for EM. (3) It augments training data with (difficult) examples, which challenges \system{} to learn ``harder'' and also reduces the amount of training data required.
%
\iffalse %%% SHORTEN
We show that the basic version of \system{} is already rather powerful: it outperforms a previous state-of-the-art (SOTA) EM solutions by an average of 3.49\% in F1 score on benchmark datasets. % up to 19%
The optimizations further boost the improvement to 6\% in average and up to 25\% in specific tasks.
% After the optimizations are applied, \system{} outperforms previous SOTA by 6\% in average
% and by up to 25\%
Furthermore, \system\ can achieve the same performance with as little as half the training data. 
While \system{}'s primary functionality is on matching, we also developed an advanced blocking method based on sentence embeddings to develop an end-to-end EM system.
\fi %%%%%% END
Figure~\ref{fig:pipeline} depicts \system{} in the overall architecture of a complete EM workflow. 

% \small The pipeline for EM. The Blocker quickly identifies a set of candidate matching pairs
% and the Matcher verifies whether each candidate pair is a real match.
% \system\ (like other learning-based EM solutions) trains the Matcher offline on a sample of labeled
% subset of candidate pairs.

There are 9 candidate pairs of entries to consider for matching in total in Figure~\ref{fig:EMExample}. The blocking heuristic that matching entries must have one word in common in the \textbf{title} will reduce
the number of pairs to only 3: the first entry on the left with the first entry on the right and so on. Perhaps more surprisingly, even though the 3 pairs are highly similar and look like matches, only the first and last pair of entries are true matches. 
Our system, \system{}, is able to discern the nuances in the 3 pairs to make the correct conclusion for every pair while some state-of-the-art systems are unable to do so.

The example illustrates the power of language understanding given by 
\system's pre-trained language model. It understands that 
%\yuliang{I removed this ``likelier'' since it is more certain.} % is likelier to understand that
\textsl{instant immersion spanish deluxe 2.0} is the same as 
\textsl{instant immers spanish dlux 2} in the context of software products even though they are syntactically different. 
%that \textsl{academic} is essentially same as \textsl{student/staff} in %the context of software products in the first candidate pair, even though %they are syntactically different.
%Furthermore, the attention mechanism of language models makes it likelier for \system{} to learn what are the parts of values that are critical for making the right conclusions and consequently, this also makes \system{} more robust to dirty data. 
%When the data is a long text, the text is summarized to help \system{}
%focus on the essential information. 
Furthermore, one can explicitly emphasize that certain parts of a value are more useful for deciding matching decisions. 
For books, the domain knowledge that the grade level or edition is important
for matching books can be made explicit to \system, simply by placing tags around the grade/edition values.
Hence, for the second candidate pair, 
even though the titles are highly similar (i.e., they overlap in many words), \system{} is able to focus on the grade/edition information when making the matching decision.
The third candidate pair shows the power of language understanding for the opposite situation.
%A similar phenomenon happens with the third candidate pair of entries. 
Even though the entries look dissimilar
%and the essential information (i.e., manf./modelno) is found under different attributes of the entries, 
\system{} is able to attend to the right parts of a value (i.e.,
the manf./modelno under different attributes) and also understand the semantics of the model number to make the right decision. 

%%%%%%%%%%%%%%%%

\smallskip
\noindent
{\bf Contributions~} In summary, the following are our contributions:
\begin{itemize}\parskip=0pt
\item We present \system, a novel EM solution
based on pre-trained language models (LMs) such as BERT. 
We fine-tune and cast EM as a sequence-pair classification problem to leverage such models with a simple architecture.
To our knowledge, \system\ is one of the first EM solutions that leverage pre-trained Transformer-based 
LMs\footnote{\small There is a concurrent work~\cite{brunner2020entity} which applies a similar idea.}
to provide deeper language understanding for EM.
% To the best of our knowledge, \system\ is the first EM solution 
% that leverages pre-trained Transformer-based LMs, which are powerful LMs 
% that have been shown to provide deeper language understanding. \yuliang{TODO: change this statement.s}

%\item
%We present a simple yet effective %pre-processing method for serializing 
%data entries into text sequences so that EM %can be solved as a sequence-pair %classification problem
%by fine-tuning the LMs. 

\item We also developed 
three optimization techniques to further improve \system's matching capability through injecting
domain knowledge, summarizing long strings, and augmenting training
data with (difficult) examples.
%:
%(1) by allowing domain knowledge to %be exploited through emphasizing %important
%pieces of input that may be of %interest when making matching %decisions. (2) by summarizing long %strings to retain only the most useful information, and (3) by augmenting the training data with (difficult) examples. 
The first two techniques help \system{} focus on the right information for making matching decisions. The last technique, data augmentation, is adapted from~\cite{miao2020snippext} for EM to
help \system{} learn ``harder'' to understand the data invariance properties that may exist but are beyond the provided labeled examples and also, reduce the amount of training data required.

%These first two techniques enable \system\ to pay attention to the right information when making entity matching decisions and (1) is useful for injecting domain knowledge while (2)
%is also useful for ensuring that the 
%the resulting length of the serialized %sequence is still within the hard limit %given by pre-trained LMs. The last technique %helps \system{} understand the data %invariance properties that may exists but %are beyond the provided labeled examples.

\item We evaluated the effectiveness of \system\ on three benchmark datasets: 
the Entity Resolution benchmark~\cite{kopcke2010evaluation},
the Magellan dataset~\cite{Konda:2016:Magellan}, and the WDC product matching dataset~\cite{Primpeli:2019:WDC}
of various sizes and domains. Our experimental results show that \system\
consistently outperforms the previous SOTA EM solutions in all datasets and
by up to \rev{31\%} in F1 scores. 
%\system\ is also more robust to noisy data as confirmed by our error analysis.
%\wctan{are we going ot have an error analysis?}
Furthermore, \system{} consistently performs better on dirty data and is more label efficient:
it achieves the same or higher previous SOTA accuracy using less than half the labeled data. 
%We open-source our implementation of \system\ at \url{github.com}
%\wctan{we should open source this?} \yuliang{yes but I don't think we %have enough time to 
%make the codebase ready by the deadline.}
\item We applied \system\ to 
a real-world large-scale matching task on two company datasets, containing 789K and 412K entries respectively.
%\yuliang{why don't we mention employer matching here? I think it sounds more practical.}
To deploy an end-to-to EM pipeline efficiently,
we developed an advanced blocking technique
%based on sequence embeddings and vector similarity search 
to help reduce the number of pairs to consider for \system. 
\system{} obtains high accuracy, 96.5\% F1 on a holdout dataset.
The blocking phase also helped speed up the end-to-end EM deployment significantly, by up to 3.8 times, compared to naive blocking techniques.
%\wctan{I am not sure the last stmt is a strong one. It seems to beg the question of using and comparing with other blocking techniques}
\item Finally, we open-source \system{} at \small{ \url{https://github.com/megagonlabs/ditto}}.
%\wctan{added this}

%Our case study shows that \system\ leads to %immediate performance gain.
%The pipeline features an advanced blocking %technique based on 
%the SOTA sequence %embeddings~\cite{reimers2019sentence}
%and vector similarity search accelerating the %overall pipeline by 3.8x.
% Are more robust to noise (Dirty vs. Structured)
% Are more label efficient as they already outperformed the previous SOTA when given only 1/Y of the training data.

% Proposed pre-processing methods that 
% serialize structured data into sequence data while keeping the structural information,
% inject domain knowledge to the training/testing data via Name Entity Recognition

% Additionally, we applied the SOTA Data Augmentation technique (MixDA) to make the training data more challenging which results in more robust and accurate models.
% DEL / SWAP now
% Multiple levels of operators (column-level, token-level)
% We conducted experiments on 13 deepmatcher datasets and the WDC product datasets of various sizes and domains. Our experimental results show that our models 
% Outperformed the previous SOTA solution by up to X%, 
% Are more robust to noise (Dirty vs. Structured)
% Are more label efficient as they already outperformed the previous SOTA when given only 1/Y of the training data.
% Explanation of why BERT/LMs work well on structured data
% Attention
% Error Analysis
% A real case study on the Glassdoor dataset
% Heuristics-based blocking  =>  labeled
% Comparison with DeepMatcher
\end{itemize}

\noindent
{\bf Outline~} 
Section \ref{sec:architecture} overviews \system\ and pre-trained LMs. 
Section \ref{sec:optimizations} describes how we optimize \system\
with domain knowledge, summarization, and data augmentation.
Our experimental results are described in Section \ref{sec:experiments} 
and the case study is presented in Section \ref{sec:casestudy}. 
We discuss related work in Section \ref{sec:related} and conclude in Section \ref{sec:conclusion}.

%%%% END OLD
\newtheorem{problem}{Problem}

\newcommand{\attr}{\mathsf{attr}}
\newcommand{\val}{\mathsf{val}}
\newcommand{\serialize}{\mathsf{serialize}}
\newcommand{\augment}{\mathsf{augment}}

\section{Background and Architecture}\label{sec:architecture}

We present the main concepts behind EM and  
provide some background on pre-trained LMs before we describe
%first review the problem definition %of EM in the context of 
%a complete entity matching workflow. %Next, we review pre-trained LMs and 
how we fine-tune the LMs on EM datasets to train EM models.
We also present a simple method for reducing EM to a sequence-pair classification problem so that pre-trained LMs can be used for solving the EM problem.
%serializing data entries for \system{} for matching.
%LMs, which constitutes
%the foundation for ingesting data entries %into \system{} for matching.

% \yuliang{Need a system architecture.}

%\subsection{A complete EM workflow}
%\wctan{I wonder if readers will find this too long}
\smallskip
\noindent
{\bf Notations~}
%\system{}
% {\color{red} An EM system} 
\system's EM pipeline takes as input two collections $D$ and $D'$ of data entries 
(e.g., rows of relational tables, XML documents, JSON files, text paragraphs)
and outputs a set $M \subseteq D \times D'$ of pairs where 
each pair  $(e, e') \in M$ is thought to represent 
the same real-world entity (e.g., person, company, laptop, etc.).
%\yoshi{$i$ is also used below. I think we can simply remove $i$ as we describe $e$ and $e'$ in the following examples.} \yuliang{good point. fixed}
A data entry $e$ is a set of key-value pairs
$e = \{(\attr_i, \val_i)\}_{1 \leq i \leq k}$ where $\attr_i$ is the attribute name
and $\val_i$ is the attribute's value represented as text. Note that our definition
of data entries is general enough to capture both structured and 
semi-structured data such as JSON files.

As described earlier, an end-to-end EM system consists of a {\em blocker} and a 
{\em matcher}.
The goal of the blocking phase is to quickly identify a small subset of $D \times D'$ of candidate pairs of high recall (i.e., a high proportion of actual matching pairs are that subset). 
%Blocking is necessary when testing all $|D \times D'|$ pairs is %infeasible. 
%For instance, when matching two tables of employer entries with address %information,
%one effective blocking method is to find pairs that
%%that have the same \textsf{zipcode}.
%Although this method would produce many false positives,
%it is fast and can effectively reduce the number of candidate pairs.
The goal of a matcher (i.e., \system{}) is to 
accurately predict, given a pair of entries, whether they refer to the same real-world entity. 
%\system{} is a matcher although we also describe an advanced %blocking technique that we have developed in %Section~\ref{sec:casestudy}.
%\yoshi{I know we are not sure at this moment. But, it should be better to clarify whether Ditto focues on Matching or can be used for both.} 
%optimizes the \emph{matching} phase of EM. 
%During the \emph{matching} phase, each candidate %pair is examined to determine
%whether they are real match or not. Our goal is to %develop
\iffalse %%%
A matcher has a \emph{matching function} $F: D \times D' \mapsto \{0, 1\}$ where, in the ideal case,
$F(e, e') = 1$ if $(e, e') \in C$ is a real match.  Otherwise, $F(e, e') = 0$.
The matching function $F$ can be rules (e.g., keyword-based matching with TF-IDF) or it can be a model
learned from training data. In the latter case, a training set $T$ is sampled from $C$
and manually labeled. We then train $F$ as a binary classification model to predict whether a candidate pair matches or not.
%separate the matched pairs
%with the no-matched pairs. 
The goal is to optimize the F1 score 
%(the harmonic mean of precision and recall) 
of the learned model $F$ on a holdout test set $T'$, which is disjoint from $T$ and is constructed the same way as $T$.
The EM pipeline is depicted in Figure \ref{fig:pipeline}.
\fi %%%

% \begin{figure}[!ht]
% \centering
% \includegraphics[width=0.5\textwidth]{pipeline.pdf}
% \caption{\small The pipeline for EM. The Blocker quickly identifies a set of candidate matching pairs
% and the Matcher verifies whether each candidate pair is a real match.
% \system\ (like other learning-based EM solutions) trains the Matcher offline on a sample of labeled
% subset of candidate pairs.}
% \label{fig:pipeline}
% \end{figure}

\subsection{Pre-trained language models}

% \yoshi{@Yuliang, I moved the first two paragraphs from the subsection ``Fine-tuning...'' Please check}
Unlike prior learning-based EM solutions that rely on word embeddings and customized RNN architectures to train the matching model (See Section \ref{sec:related} for a detailed summary), \system\ trains the matching models by fine-tuning pre-trained LMs in a simpler architecture.

Pre-trained LMs such as BERT~\cite{Devlin:2019:BERT}
and GPT-2~\cite{gpt2} have demonstrated good performance 
on a wide range of NLP tasks. They are typically deep neural networks 
with multiple Transformer layers~\cite{vaswani2017attention}, typically 12 or 24 layers, pre-trained on large text corpora such as Wikipedia articles in an unsupervised manner.
During pre-training, the model is self-trained to perform auxiliary tasks such as 
missing token and next-sentence prediction. Studies~\cite{Clark:2019:WhatDoesBERTLookAt,Tenney:2019:BertRediscovers} 
have shown that the shallow layers capture lexical meaning while the deeper layers 
capture syntactic and semantic meanings of the input sequence after pre-training. 

%\yuliang{This paragraph is not reading very well. I rephrased a bit and I can get back to it.}
%\wctan{reworded}
A specific strength of pre-trained LMs is that
it learns the semantics of words better than 
conventional word embedding techniques such as word2vec, GloVe, or FastText. 
This is largely because 
the Transformer architecture calculates token embeddings 
from all the tokens in the input sequence and thus, the 
embeddings it generates are {\em highly-contextualized} and captures the 
semantic and contextual understanding of the words. 
Consequently, such embeddings can capture polysemy, i.e.,  discern that the same word may have different meanings in different phrases. For example, the word {\em Sharp} has different meanings in \textsl{``Sharp resolution''} versus \textsl{``Sharp TV''}. 
Pre-trained LMs will embed ``{\em Sharp}'' differently depending on the context
while traditional word embedding techniques such as FastText always produce the same
vector independent of the context.
% Traditional word embedding techniques such as FastText will 
% produce the same embedding for the word ``{\em Sharp}'' in both contexts. 
% This is not the case for pre-trained LMs. 
Such models can also understand the opposite, i.e., that different words may have the same meaning. For example, the words \textsl{immersion} and \textsl{immers} (respectively, (\textsl{deluxe}, \textsl{dlux}) and
(2.0, 2)) are likely the same given their respective contexts.
%as we have illustrated in our running example.
%
Thus, such language understanding capability of pre-trained LMs can improve the EM performance.

%%%

\subsection{Fine-tuning pre-trained language models}
A pre-trained LM can be fine-tuned with task-specific training data so that it becomes better at performing that task. Here,
we fine-tune a pre-trained LM for the EM task with a labeled training dataset consisting of positive and negative pairs of matching and non-matching entries
as follows:
\vspace{-1mm}
\begin{enumerate}\parskip=0pt\itemsep=0pt
\item Add task-specific layers after the final layer of the LM.
For EM, we add a simple fully connected layer 
and a softmax output layer for binary classification.
\item Initialize the modified network with parameters from the pre-trained LM.
\item Train the modified network on the training set until it converges.
\end{enumerate}
\vspace{-1mm}

The result is a model fine-tuned for the EM task. 
\rev{See Appendix \ref{sec:lm} for the model architecture.}
\rev{In \system{}, we fine-tune  
the popular 12-layer BERT model~\cite{Devlin:2019:BERT}, RoBERTa~\cite{liu2019roberta}, 
and a 6-layer smaller but faster variant DistilBERT~\cite{sanh2019distilbert}.
However, our proposed techniques are independent of the choice of pre-trained LMs and
\system\ can potentially perform even better with larger pre-trained LMs.} 
%\wctan{modified}
%\yuliang{good.}
% We illustrate the model architecture in Figure \ref{fig:lm}. 
The pair of data entries is serialized (see next section) as input to the LM and 
the output is a match or no-match decision. 
\system's architecture is much simpler when compared to many state-of-the-art EM solutions 
today~\cite{Mudgal:2018:DeepMatcher,Ebraheem:2018:DeepER}. Even though the bulk of the ``work'' is simply off-loaded to pre-trained LMs, we show that this simple scheme works surprisingly well in our experiments. 

% \begin{figure}[t]
%     \centering
%     \includegraphics[width=0.49\textwidth]{lm.pdf}
%     \caption{\small \system's model architecture. \system\ serializes the two entries as one sequence
%     and feeds it to the model as input. The model consists of (1) token embeddings and 
%     Transformer layers~\cite{wolf2019transformers} from a pre-trained language model (e.g., BERT) and 
%     (2) task-specific layers (linear followed by softmax). Conceptually, the \textsf{[CLS]} token 
%     ``summarizes'' all the contextual information needed for matching as a contextualized embedding 
%     vector $E'_{\mathsf{[CLS]}}$ which the task-specific layers take as input for classification.}
%     \label{fig:lm}
%     \vspace{-4mm}
% \end{figure}

\subsection{Serializing the data entries for Ditto}
Since LMs take token sequences (i.e., text) as input, a key challenge is
to convert the candidate pairs into token sequences so that they can be meaningfully ingested by \system.

\system\ serializes data entries as follows:
for each data entry $e = \{(\attr_i, \val_i)\}_{1 \leq i \leq k}$, we let
$$\serialize(e) ::= \mathsf{[COL] } \ \attr_1 \ \mathsf{ [VAL] } \ \val_1 \dots \mathsf{ [COL] } \ \attr_k \ \mathsf{ [VAL] } \ \val_k,$$
where $\mathsf{[COL]}$ and $\mathsf{[VAL]}$ are special tokens for indicating the start of
attribute names and values respectively. For example,the first entry of the second table is serialized as:

%\yuliang{added this example.}
\smallskip
\noindent
\begin{small}%\begin{center}
\textsf{[COL] title [VAL] instant immers spanish dlux 2}
\textsf{[COL] manf./modelno [VAL] NULL [COL] price [VAL] 36.11}
%\end{center}
\end{small}
\smallskip

To serialize a candidate pair $(e, e')$, we let
$$\serialize(e, e') ::= \mathsf{[CLS] } \ \serialize(e) \ \mathsf{ [SEP] } \ \serialize(e') \ \mathsf{[SEP]}, $$
where $\mathsf{[SEP]}$ is the special token separating the two sequences and
$\mathsf{[CLS]}$ is the special token necessary for BERT to encode the sequence pair 
into a 768-dimensional vector which will be fed into the fully connected layers for classification.

\smallskip
\noindent
\textbf{Other serialization schemes~} There are different ways to serialize data entries so that LMs can treat the input as a sequence classification problem.
%The serialization method for casting EM into sequence classification is not unique. 
For example, one can also omit
the special tokens
%Other possible methods include the variants of not adding the special 
``$\mathsf{[COL]}$'' and/or ``$\mathsf{[VAL]}$'',
or exclude attribute names $\attr_i$ during serialization. 
We found that including the special tokens 
to retain the structure of the input does not hurt the performance in general and excluding the attribute names tend to help only when the attribute names do not contain useful information (e.g., names such as attr1, attr2, ...) or when the entries contain only one column. 
%These are all valid options and sometimes can perform better than the above base form
%especially when there is only 1 attribute in $e$ and $e'$ or when the attribute names %do not carry useful information.
A more rigorous study on this matter is left for future work. 

%\wctan{add references to system if we have not referenced them before this point}
\smallskip
\noindent
\textbf{Heterogeneous schemas~}
As shown, the serialization method of \system{} does not require data entries to adhere to the same schema. It also does not require that the attributes of data entries to be matched prior to executing the matcher, which is a sharp contrast to other EM systems such as DeepER~\cite{Ebraheem:2018:DeepER} or DeepMatcher\footnote{\small In DeepMatcher, the requirement that both entries have the same schema can be removed by treating the values in all columns as one value under one attribute.}~\cite{Mudgal:2018:DeepMatcher}. 
Furthermore, \system{} can also ingest and match hierarchically structured data entries 
by serializing nested attribute-value pairs with special start and end tokens 
(much like Lisp or XML-style parentheses structure).

%However, we found that the base form generally performs better or equally well compared %to the other variants.
%This indicates that (1) the attribute names usually contain useful information and (2) %keeping the input more structured and less ambiguous helps. \yoshi{The paragraph looks %good to me except the last statement. I feel like we may want to move the statement to %Discussion section with results that supports the claim. Probably, we can simply add a %sentence ``We will discuss it in detail in Section xx''. or something like that.}
%% \yuliang{I think we can drop the statement. }
%\yuliang{can someone help shorten this paragraph?}

% \smallskip
% \noindent
% \textbf{Sequence length. }

% \yuliang{shall we say that distorting the grammar is okay?}
% \yoshi{We could emphasize that Ditto works robustly with such a simple filtering. Technically speaking, DeepMatcher showed a improvement by their ``information extraction'' module, though.}

\section{Optimizations in Ditto}
\label{sec:optimizations}

As we will describe in Section~\ref{sec:experiments}, the basic version of
\system{}, which leverages only the pre-trained LM, is already outperforming the SOTA on average. Here, 
we describe three further optimization techniques that will facilitate and challenge \system{} to learn ``harder'', and consequently make better matching decisions.

\newcommand{\recognizer}{\mathsf{recognizer}}
\newcommand{\rewriter}{\mathsf{rewriter}}
\newcommand{\type}{\mathsf{type}}

\subsection{Leveraging Domain Knowledge}\label{sec:domain}

Our first optimization allows domain knowledge to be injected into \system{}
through {\em pre-processing} the input sequences (i.e., serialized data entries) to 
emphasize what pieces of information are potentially important. 
%to the matching models via 
%\emph{pre-processing the input %sequences}. 
This follows the intuition that
when human workers make a matching/non-matching decision on two data entries, they typically look for 
spans of text that contain key information before making the final decision.
%follow certain patterns
%when comparing two entities of each %type.
%The human workers might (1) focus on %certain spans of text to make the %matching decision
%or (2) decide that two spans are the %same although they are slightly %different.
Even though we can also
train deep learning EM solutions to learn such knowledge, we will require a significant amount of training data to do so. As we will describe, this pre-processing step on the input sequences is lightweight 
and yet can yield significant improvements. 
Our experiment results show that with less than 5\% of additional training time,
we can improve the model's performance by up to 8\%.
%\yuliang{added this.}

% \subsection{Domain knowledge injection}
%We propose injecting the domain %knowledge to the matching models via 
%\emph{pre-processing the input %sequences}. 
%In \system, we consider two main categories of domain knowledge.
There are two main types of domain knowledge that we can provide \system.

\smallskip
\noindent
{\bf Span Typing~} 
The type of a span of tokens is one kind of domain knowledge that can be provided to \system. Product id, street number, publisher are examples of span types. Span types help \system{} avoid mismatches. With span types, for example, \system{} is likelier to avoid
matching a street number with a year or a product id.

%consists of 
%identifying spans of important types %for matching a certain type of %entities.
%For example, when checking whether two %company records represent the same %employer,
%human workers would examine whether %their phone numbers are identical and
%check their street numbers are the same.

Table \ref{tab:spans} summarizes the main span types that human workers
would focus on when matching three types of entities in our benchmark datasets.

\setlength{\tabcolsep}{3pt}
\begin{table}[!ht]
\vspace{-2mm}
\small
\centering
\caption{\small Main span types for matching entities in our benchmark datasets.}
\vspace{-4mm}
\label{tab:spans}
\begin{tabular}{ll}
\toprule
\textbf{Entity Type}             & \textbf{Types of Important Spans} \\ \midrule
Publications, Movies, Music & Persons (e.g., Authors), Year, Publisher \\
Organizations, Employers   & Last 4-digit of phone, Street number \\ 
Products                  & Product ID, Brand, Configurations (num.) \\ \bottomrule
\end{tabular}
    \vspace{-4mm}
\end{table}

The developer specifies a $\recognizer$ to type
spans of tokens from attribute values. 
The $\recognizer$ takes a text string $v$ as input
and returns a list $\recognizer(v) = \{(s_i, t_i, \type_i)\}_{i \geq 1}$
of start/end positions of the span in $v$ and the corresponding type of the span.
\system's current implementation leverages an open-source Named-Entity Recognition (NER) model~\cite{spacy:ner} %\footnote{\small \url{https://spacy.io/api/entityrecognizer}} 
to 
identify known types such as persons, dates, or organizations and
use regular expressions to identify specific types such as product IDs, last 4 digits of phone numbers, etc.

After the types are recognized,
the original text $v$ is replaced by
a new text where special tokens are inserted to reflect the types of the spans.
For example,  a phone number ``\textsf{(866) 246-6453}''
may be replaced with ``\textsf{( 866 ) 246 - [LAST] 6453 [/LAST]}'' where \textsf{[LAST]}/\textsf{[/LAST]} indicates
the start/end of the last 4 digits and additional spaces are also added because of tokenization.
% \wctan{don't we need an end token in general? need to explain}
In our implementation, when we are sure that the span type has only one token or the NER model is inaccurate in determining the end position, we drop the end indicator and keep only the start indicator token.
%\yuliang{added the end token.}

Intuitively, these newly added special tokens are additional signals to the
self-attention mechanism that already exists in pre-trained LMs, such as BERT. If two spans have the same type, then \system{} picks up the signal that 
they are likelier to be the same and hence, they 
are aligned together for matching.
%\wctan{stopped here}
In the above example, 
%\vspace{-0.75mm}

\vspace{-3mm}
\begin{small}
$$`` .. \textsf{246- } \textsf{[LAST] 6453 [/LAST] .. [SEP] } .. \textsf{ [LAST] 0000 [/LAST]} .. "
\vspace{-0.5mm}
$$\end{small}
when the model sees two encoded sequences with the \textsf{[LAST]} special tokens, 
it is likely to take the hint to align ``6453'' with ``0000'' without relying on other
patterns elsewhere in the sequence that may be harder to learn.
%\jinfeng{It is a bit hard to understand ``align'' here. Any more explanation?}

%\subsection{Span normalization} 
\smallskip
\noindent
{\bf Span Normalization~}
The second kind of domain knowledge that can be passed to \system{} rewrites syntactically different but equivalent spans into the same string. This way, they will have identical embeddings and it becomes easier for \system{} to detect that the two spans are identical.
For example, we can enforce that ``\textsf{VLDB journal}'' and ``\textsf{VLDBJ}''
are the same by writing them as \textsf{VLDBJ}. Similarly, we can enforce the general knowledge that ``\textsf{5 \%}'' vs. ``\textsf{5.00 \%}'' are equal by writing
them as ``\textsf{5.0\%}''. 
%when dealing with numbers, the word embeddings of two same values
%``\textsf{5 \%}'' vs. ``\textsf{5.00 \%}'' will treat them as ``similar'' instead of exactly %the same.
%Similarly for abbreviations, phrases like ``\textsf{VLDB journal}'' and ``\textsf{VLDBJ}'' %are encoded 
%very differently in the embedding space, while domain experts can immediately
%tell these are the same spans and they are all different from ``\textsf{PVLDB}''.

The developer specifies 
a set of rewriting rules to rewrite spans. 
The specification consists of a function that first identifies the spans of interest before it replaces them with the rewritten spans.
%\wctan{I don't understand why the recognizer has to be applied first and why we can only rewrite spans recognized by span typing. Aren't they independent?} \yuliang{rewritten.}
\system{} contains a number of 
rewriting rules for numbers, including rules that round all floating point numbers to 2 decimal places
and dropping all commas from integers (e.g., ``2,020'' $\rightarrow$ ``2020'').
For abbreviations, we allow the developers to specify a dictionary of synonym pairs
to normalize all synonym spans to be the same.

%\yuliang{Shall we mention SentencePiece? I don't want to confuse people but it is part of why we do this rewriting.}\yoshi{If we emphasize that one of the benefits is given by SentencePiece, yes. I think natural reaction from reviewers would be ``What if you use SentencePiece for DeepMatcher?'' Well, subword tokenization without self-attention should not work well (i.e., SentencePiece + RNN (DeepMatcher) should not work robustly.) So, we can logically explain the benefit of SentencePiece and self-attention at the same time?}

%\noindent
%{\bf Discussion~}
%\wctan{pls write your sentence piece discussion %here}
%\yuliang{I suggest dropping the SentencePiece %discussion.}

\subsection{Summarizing long entries}
When the value is an extremely long string, it becomes harder for the LM to 
understand what to pay attention to when matching. In addition,
one limiting factor of Transformer-based pre-trained LMs is that there is a
limit on the sequence length of the input. 
For example, the input to BERT can have at most 512 sub-word tokens.
It is thus important to summarize the serialized entries down to the maximum allowed length
while retaining the key information. 
A common practice is to truncate the sequences so that
they fit within the maximum length. However, the truncation strategy does not work well for EM
in general because the important information for matching is usually not 
at the beginning of the sequences. 

There are many ways to perform summarization~\cite{mihalcea2004textrank,radford2019language,rush2015neural}. 
In \system's current
implementation, we use a TF-IDF-based summarization technique 
that \emph{retains non-stopword tokens with the high TF-IDF scores}.
We ignore the start and end tags generated by span typing in this process and use the list of stop words from scikit-learn library\rev{~\cite{sklearn}}.
By doing so, \system\ feeds only the most informative tokens to the LM. 
We found that this technique works well in practice.
Our experiment results show that it improves the F1 score of \system\ on
a text-heavy dataset from \rev{41\% to over 93\%} and we plan to add more summarization techniques to \system's library in the future.

%\yuliang{rewrote this paragraph. Pls check.}
%\wctan{big question. won't the summarization mess up the typing information in the previous substep? we need ot explain this.} \yuliang{it ignores all the special tokens that we generated.}

\newcommand{\lm}{\mathsf{LM}}

\subsection{Augmenting training data}\label{sec:augment}
%\wctan{should we make this the first subsection?}
We describe how we apply data augmentation
to augment the training data for entity matching.

Data augmentation (DA) is a commonly used technique in computer vision for generating
additional training data from existing examples by simple transformation
such as cropping, flipping, rotation, padding, etc. 
The DA operators not only add more training data, but the augmented data also 
allows to model to learn to make predictions invariant of these transformations.

Similarly, DA can add training data that will help EM models learn ``harder''.
Although labeled examples for EM are arguably not hard to obtain,
invariance properties are very important to help make the solution 
more robust to dirty data, such as missing values (NULLs), 
values that are placed under the wrong attributes or missing some tokens. 

Next, we introduce a set of DA operators for EM
that will help train more robust models.
%in practice the EM datasets very often contain %incomplete or misplaced information
%(i.e., NULLs or values being placed at a wrong %attribute).
%In such cases, DA can make the trained models %more robust 
%by using a set of carefully chosen %transformation operators.
%We propose a set of DA operations suitable for %EM as our first contribution.

\smallskip
\noindent
{\bf Augmentation operators for EM~} The proposed DA operators are summarized in Table \ref{tab:da}.
If $s$ is a serialized pair of data entries with a match or no-match label $l$, then
an augmented example is a pair $(s',l)$, where $s'$ is obtained
by applying an operator $o$ on $s$ and $s'$ has the same label $l$ as before. 
%Formally, given a serialized data entry pair $s = \serialize(e, e')$
%and an operator $o$, we generate an %augmented example denoted by
%$s' = \augment(s, o)$ which is a randomized %output of applying $o$ on $s$.

\setlength{\tabcolsep}{4pt}
\begin{table}[!ht]
\small
\centering
\caption{\small Data augmentation operators in \system. The operators are 3 different levels:
span-level, attribute-level, and entry-level. 
All samplings are done uniformly at random.} \label{tab:da}
\vspace{-3mm}
\begin{tabular}{cc}
\toprule
\textbf{Operator}      & \textbf{Explanation}  \\ \midrule
span\_del     & Delete a randomly sampled span of tokens                      \\
span\_shuffle & Randomly sample a span and shuffle the tokens' order \\
attr\_del     & Delete a randomly chosen attribute and its value              \\
attr\_shuffle & Randomly shuffle the orders of all attributes                 \\
entry\_swap   & Swap the order of the two data entries $e$ and $e'$              \\ \bottomrule
\end{tabular}
    \vspace{-1.5mm}
\end{table}

The operators are divided into 3 categories.
The first category consists of span-level operators, such as
span\_del and span\_shuffle. 
These two operators are used in NLP tasks~\cite{wei2019eda,miao2020snippext} and 
shown to be effective for text classification.
For span\_del, we
randomly delete from $s$ 
a span of tokens of length at most 4 without special tokens (e.g., \textsf{[SEP], [COL], [VAL]}).
For span\_shuffle, we sample a span of length at most 4
and randomly shuffle the order of its tokens.

These two operators are motivated by the observation that making a match/no-match decision 
can sometimes be ``too easy'' when the candidate pair of data entries contain multiple spans of text 
supporting the decision. For example, suppose our negative examples for matching company data in the existing training data is similar to what is shown below.

\vspace{-1mm}
\begin{small}
\begin{center}
\textsf{[CLS] $\dots$ [VAL] Google LLC $\dots$ [VAL] (866) 246-6453 [SEP] $\dots$ } \\
\textsf{[VAL] Alphabet inc $\dots$ [VAL] (650) 253-0000 [SEP]}
\end{center}
\end{small}
\vspace{-1mm}
The model may learn to predict ``no-match'' based on
the phone number alone, which is insufficient in general.
On the other hand, by corrupting parts of the input sequence (e.g., dropping phone numbers), 
DA forces the model to learn beyond that,
by leveraging the remaining signals, such as the company name, to predict ``no-match''.

%both the company names and phone numbers are useful signals for making the matching decision on company data such as the two companies shown above.
%When we train the model, if the model has already learned to match by the phone numbers,
%because there is no gradient descend,
%this unaugmented example does not cause the model to further learn to match by company name.
%On the other hand, by corrupting parts of the input sequence (e.g., dropping a phone number), 
%DA forces the model to learn to leverage the remaining signals to match the pair.

The second category of operators is attribute-level operators: attr\_del and attr\_shuffle.
The operator attr\_del randomly deletes an attribute (both name and value) and
attr\_shuffle randomly shuffles the order of the attributes of both data entries.
The motivation for attr\_del is similar to span\_del and span\_shuffle but 
it gets rid of an attribute entirely.
The attr\_shuffle operator allows the model to learn the property that
the matching decision should be independent of the ordering of attributes in the sequence.
% This invariance property is especially useful for datasets with missing attributes.
% \yuliang{This claim might not be true.}

The last operator, entry\_swap, swaps the order of the pair $(e, e')$ 
with probability $1/2$. This teaches the model to make symmetric decisions (i.e., $F(e, e') = F(e', e)$)
and helps double the size of the training set if both input tables
are from the same data source.

% \yuliang{TODO: if we have space, we can write about composing the operators}
\smallskip
\noindent
\textbf{MixDA: interpolating the augmented data }
Unlike DA operators for images which almost always preserve the image labels, 
the operators for EM can distort the input sequence
so much that the label becomes incorrect. For example, the attr\_del operator
may drop the company name entirely and the remaining
attributes may contain no useful signals to distinguish the two entries.

\begin{figure}[!ht]
    \vspace{-3mm}
    \centering
    \includegraphics[width=0.42\textwidth]{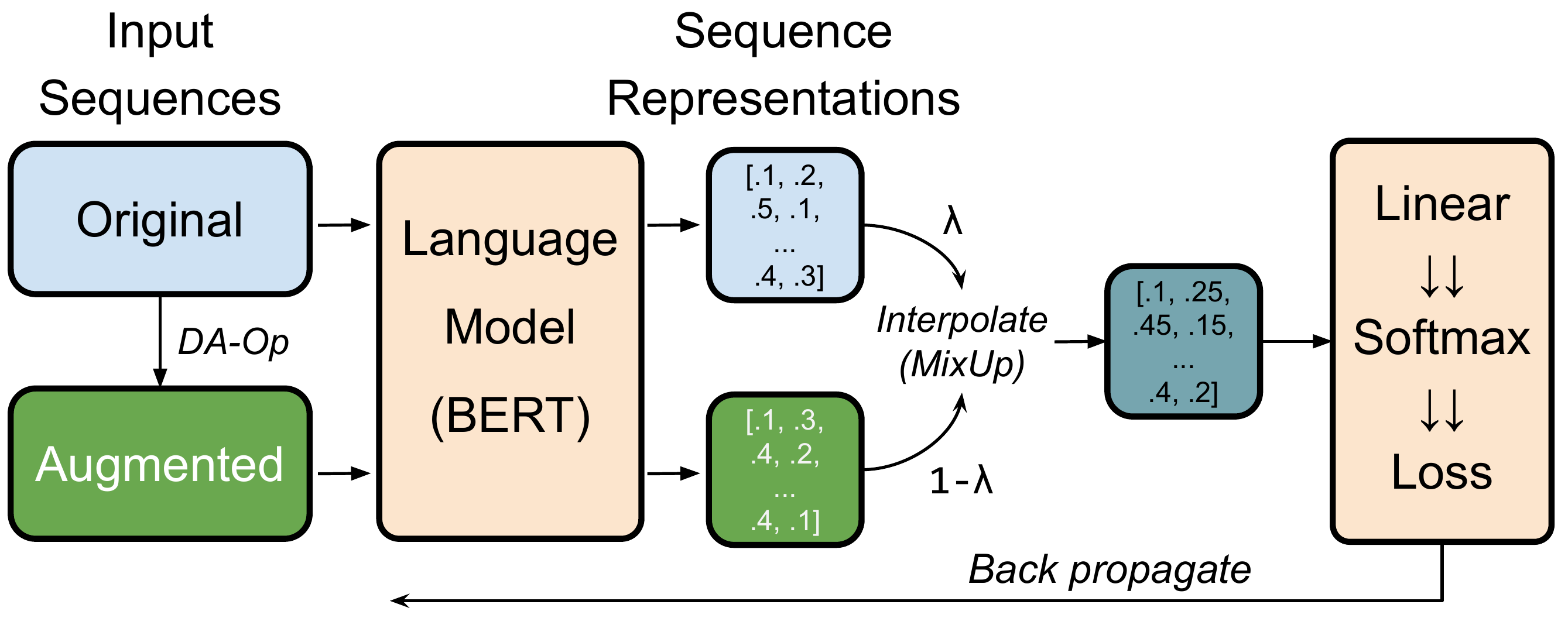}
    \vspace{-4mm}
    \caption{\small Data augmentation with MixDA.}
    % \caption{\small Data augmentation with MixDA. To apply MixDA, we first transform the example with a DA
    % operator and pass it to the LM. We then interpolate the representations of the original and the augmented examples. Finally, we feed the interpolation to the rest of the NN and back-propagate.}
    \label{fig:mixda}
    \vspace{-4mm}
\end{figure}

To address this issue, \system\ applies MixDA, a recently proposed
data augmentation technique for NLP tasks~\cite{miao2020snippext} illustrated in Figure \ref{fig:mixda}.
Instead of using the augmented example directly, 
MixDA computes a convex interpolation of the original example with the augmented examples. Hence,
the interpolated example is somewhere in between, i.e., it is a ``partial'' augmentation of the original example and
this interpolated example is expected to be less distorted than the augmented one. 
% \jinfeng{Any example in real dataset?}
% \wctan{agree it will be good to show an interpolated example}

The idea of interpolating two examples is originally proposed for computer vision tasks~\cite{zhang2017mixup}.
%where images can be directly interpolated. 
For EM or text data, since we cannot directly interpolate sequences,
MixDA interpolates their representations by the language model instead.
% We omit the technical details and refer the interested readers to \cite{miao2020snippext}.
In practice, augmentation with MixDA slows the training time because the LM is called twice.
However, the prediction time is not affected since the DA operators are only applied to training data.
% \wctan{i think we can remove this? and say we refer the interested reader to [28] for more information.}
% \yuliang{added the following paragraph for MixDA.}
\rev{
Formally, given an operator $o$ (e.g., span deletion) and an original example $s$,
to apply $o$ on $s$ with MixDA (as Figure \ref{fig:mixda} illustrates),
\begin{enumerate}\parskip=0pt\itemsep=0pt
\item Randomly sample $\lambda$ from a Beta distribution $\lambda \sim \text{Beta}(\alpha, \alpha)$ with a hyper-parameter
$\alpha \in [0, 1]$ (e.g., 0.8 in our experiments);
\item Denote by $\lm(s)$ the LM representation of a sequence $s$.
Let 
\vspace{-1mm}
$$\lm(s'') = \lambda \cdot \lm(s) + (1 - \lambda) \cdot \lm(\augment(s, o)).
\vspace{-1mm}$$
Namely, $\lm(s'')$ is the convex interpolation between the LM outputs of $s$ and the augmented $s'= \augment(s, o)$;
%\yoshi{Should $\augment(s', o)$ be $\augment(s, o)$? I thought $s' = \augment(s, o)$}
%\yuliang{you're right.}
\item Train the model by feeding $\lm(s'')$ to the rest of the network and back-propagate.
Back-propagation updates both the $\lm$ and linear layer's parameters.
\end{enumerate}
\vspace{-1mm}}

% Intuitively, the Beta distribution generates a $\lambda$ close to either 0 or 1
% so $\lm(s'')$ is closer to the representation of a real sequence.
% The augmented sequence $s''$ is not actually generated thus can be viewed as a virtual sequence.
% However, its representation $\lm(s'')$ is sufficient for training the network.
% \wctan{but keep this part onwards?}
% Note that in practice, augmentation with MixDA causes slower training time because the LM is called twice,
% but the prediction time is not affected since the DA operators are only applied to training data.
% \definecolor{ForestGreen}{rgb}{0.13, 0.55, 0.13}
\definecolor{ForestGreen}{rgb}{0.0, 0.66, 0.47}
\definecolor{RubineRed}{rgb}{1.0, 0.0, 0.31}

\newcommand{\green}[1]{{\textcolor{ForestGreen}{{#1}}}}
\newcommand{\red}[1]{{\textcolor{RubineRed}{{#1}}}}

\vspace{-1mm}
\section{Experiments}\label{sec:experiments}
We present the experiment results on benchmark datasets for EM:
the ER Benchmark datasets~\cite{kopcke2010evaluation},
the Magellan datasets~\cite{Konda:2016:Magellan} and the WDC product data corpus~\cite{Primpeli:2019:WDC}.
\system\ achieves new SOTA results on all these datasets and outperforms the previous best results
by up to \rev{31\%} in F1 score.
The results show that \system\ is more robust to dirty data and performs well when the training set is small.
\system\ is also more label-efficient as it achieves the previous SOTA results using 
only 1/2 or less of the training data across multiple subsets of the WDC corpus.
Our ablation analysis shows that 
(1) using pre-trained LMs contributes to over 50\% of \system's performance gain and
(2) all 3 optimizations, domain knowledge (DK), summarization (SU) and data augmentation (DA),
are effective.
For example, SU improves the performance on a text-heavy dataset by \rev{52\%},
DK leads to \rev{1.2\%} average improvement on the ER-Magellan datasets and
DA improves on the WDC datasets by 2.53\% on average. 
\rev{In addition, we show in Appendix \ref{sec:time} that although \system\ leverages deeper
neural nets, its training and prediction time is comparable to the SOTA EM systems.}

% \yuliang{need to see how the summarization fits in.}
% For example, DK improves the average performance on the ER-Magellan datasets by 1.98\%
% and DA improves 
% \jinfeng{what is the average improvement of DK and DA?}

% \yuliang{TODO: a paragraph of the main results / findings.}

\vspace{-1mm}
\subsection{Benchmark datasets}

We experimented with all the 13 publicly available datasets used for evaluating 
DeepMatcher~\cite{Mudgal:2018:DeepMatcher}. These datasets
are from the ER Benchmark datasets~\cite{kopcke2010evaluation} and 
the Magellan data repository~\cite{magellandata}.
We summarize the datasets in Table \ref{tab:magellan-dataset}
and refer to them as ER-Magellan.
These datasets are for training and evaluating matching models for various domains
including products, publications, and businesses.
Each dataset consists of candidate pairs from two structured tables of entity records of the same schema. 
The pairs are sampled from the results of blocking and manually labeled.
The positive rate (i.e., the ratio of matched pairs) ranges from 9.4\% (Walmart-Amazon) to 25\% (Company).
The number of attributes ranges from 1 to 8.

Among the datasets, the Abt-Buy and Company datasets are text-heavy meaning that
at least one attributes contain long text. 
% For example, the data entries of the Company dataset
% contain only one attribute whose value is a whole article about a company.
Also, following \cite{Mudgal:2018:DeepMatcher},
we use the dirty version of the DBLP-ACM, DBLP-Scholar, iTunes-Amazon, and Walmart-Amazon datasets
to measure the robustness of the models against noise. 
These datasets are generated from the clean version by
randomly emptying attributes and appending their values
to another randomly selected attribute.

Each dataset is split into the training, validation, and test sets using the ratio of \textsf{3:1:1}.
\rev{The same split of the datasets is also used in the evaluation of other EM 
solutions~\cite{Fu:2019:End2End,Kasai:2019:LowResourceER,Mudgal:2018:DeepMatcher}.}
We list the size of each dataset in Table \ref{tab:magellan}.

\begin{table}[t]
\small
    \centering
    \caption{\small The 13 datasets divided into 4 categories of domains. 
    The datasets marked with $\dag$ are text-heavy (Textual).
    Each dataset with $*$ has an additional dirty version to test the models' robustness against
    noisy data. }
    \vspace{-3mm}
    \label{tab:magellan-dataset}
\begin{tabular}{cc}
\toprule
\textbf{Datasets}                              & \textbf{Domains}                \\ \midrule
Amazon-Google, Walmart-Amazon$^*$         & software / electronics \\
Abt-Buy$^\dag$, Beer                         & product                \\
DBLP-ACM*, DBLP-Scholar*, iTunes-Amazon* & citation / music       \\
Company$^\dag$, Fodors-Zagats                & company / restaurant   \\ \bottomrule
\end{tabular}
\vspace{-3mm}\end{table}

% WDC product matching \cite{Primpeli:2019:WDC} and Magellan EM \cite{Konda:2016:Magellan}
% perhaps cs784

The WDC product data corpus~\cite{Primpeli:2019:WDC} contains 26 million product offers and descriptions collected from e-commerce 
websites~\cite{wdc}. %\footnote{\small{\url{http://webdatacommons.org/largescaleproductcorpus/v2/index.html}}}.
The goal is to find product offer pairs that refer to the same product.
To evaluate the accuracy of product matchers, the dataset provides 4,400 manually created 
golden labels of offer pairs from 4 categories: computers, cameras, watches, and shoes.
Each category has a fixed number of 300 positive and 800 negative pairs.
For training, the dataset provides for each category 
pairs that share the same product ID such as GTINs or MPNs mined from the product's webpage.
The negative examples are created by selecting pairs that have high textual similarity but different IDs.
These labels are further reduced to different sizes to test the models' label efficiency.
We summarize the different subsets in Table \ref{tab:wdc-dataset}.
We refer to these subsets as the WDC datasets.
% \yuliang{TODO: mention that we only use titles.}

\setlength{\tabcolsep}{4.5pt}
\begin{table}[!ht]
\vspace{-3mm}
\small
\centering
\caption{\small Different subsets of the WDC product data corpus. Each subset (except Test) 
is split into a training set and a validation set with a ratio of \textsf{4:1}
\rev{according to the dataset provider~\protect\cite{Primpeli:2019:WDC}}. 
The last column shows the positive rate (\%POS) of each category in the xLarge set.
The positive rate on the test set is 27.27\% for all the categories.} \label{tab:wdc-dataset}
\vspace{-3mm}
\begin{tabular}{ccccccc}\toprule
\textbf{Categories} & \textbf{Test} & \textbf{Small}   & \textbf{Medium}  & \textbf{Large}   & \textbf{xLarge}  & \textbf{\%POS} \\ \midrule
Computers & 1,100 & 2,834 & 8,094  & 33,359  & 68,461  & 14.15\% \\
Cameras   & 1,100 & 1,886 & 5,255  & 20,036  & 42,277  & 16.98\% \\
Watches   & 1,100 & 2,255 & 6,413  & 27,027  & 61,569  & 15.05\% \\
Shoes     & 1,100 & 2,063 & 5,805  & 22,989  & 42,429  & 9.76\%  \\
All       & 4,400 & 9,038 & 25,567 & 103,411 & 214,736 & 14.10\% \\ \bottomrule
\end{tabular}
\vspace{-3mm}
\end{table}

\rev{
Each entry in this dataset has 4 attributes: \textsf{title}, \textsf{description}, 
\textsf{brand}, and \textsf{specTable}.
% \system{} uses only the \textsf{title} attribute because it contains rich product information such as brands, IDs, and price, making the rest of the attributes redundant.
Following the setting in \cite{Primpeli:2019:WDC} for DeepMatcher, 
we allow \system\ to use any subsets of attributes 
to determine the best combination.
We found in our experiments that \system\ achieves the best performance when it uses only the 
\textsf{title} attribute. We provide further justification of this choice in Appendix \ref{sec:wdc}.
}

\subsection{Implementation and experimental setup}\label{sec:setup}

We implemented \system\ in PyTorch~\cite{paszke2019pytorch} and 
the Transformers library~\cite{wolf2019transformers}.
\rev{
We currently support 4 pre-trained models:
DistilBERT~\cite{sanh2019distilbert},
BERT~\cite{Devlin:2019:BERT}, 
RoBERTa~\cite{liu2019roberta}, and XLNet~\cite{yang2019xlnet}.
We use the base uncased variant of each model in all our experiments.
We further apply the half-precision floating-point (fp16) optimization 
to accelerate the training and prediction speed.}
In all the experiments, we fix the max sequence length to be 256 and the learning rate to be 3e-5 
\rev{with a linearly decreasing learning rate schedule}.
The batch size is 32 if MixDA is used and 64 otherwise.
The training process runs a fixed number of epochs 
(10, 15, or 40 depending on the dataset size) and returns
the checkpoint with the highest F1 score on the validation set.
We conducted all experiments on a \textsf{p3.8xlarge} AWS EC2 machine 
with 4 V100 GPUs (1 GPU per run). 

\smallskip
\noindent
\textbf{Compared methods. } We compare \system\ with the SOTA EM solution DeepMatcher.
\rev{We also consider other baseline methods
including Magellan~\cite{Konda:2016:Magellan}, DeepER~\cite{Ebraheem:2018:DeepER},
and follow-up works of DeepMatcher~\cite{Fu:2019:End2End,Kasai:2019:LowResourceER}.}
We also compare with variants of \system\ without the data augmentation (DA) and/or domain knowledge (DK) 
optimization to evaluate the effectiveness of each component.
We summarize these methods below. We report the average F1 of 5 repeated runs in all the settings.
\vspace{-1mm}
\begin{itemize}\itemsep=0pt\parskip=0.5pt
\item \textbf{DeepMatcher: } DeepMatcher~\cite{Mudgal:2018:DeepMatcher} is the SOTA matching solution. 
Compared to \system, DeepMatcher customizes the RNN architecture to
aggregate the attribute values, then compares/aligns the aggregated representations of 
the attributes. DeepMatcher leverages FastText~\cite{fasttext} to train the word embeddings.
% \yuliang{TODO: maybe we need more details depending on what we want to compare.}
When reporting DeepMatcher's F1 scores, we use the numbers in \cite{Mudgal:2018:DeepMatcher}
for the ER-Magellan datasets and numbers in \cite{Primpeli:2019:WDC} for the WDC datasets.
We also reproduced those results using the open-sourced implementation.
\item \textbf{DeepMatcher+: } Follow-up work \cite{Kasai:2019:LowResourceER} slightly outperforms DeepMatcher in the DBLP-ACM dataset
and \cite{Fu:2019:End2End} achieves better F1 in the Walmart-Amazon and Amazon-Google datasets.
According to \cite{Mudgal:2018:DeepMatcher}, the Magellan system 
(\cite{Konda:2016:Magellan}, based on classical ML models) outperforms DeepMatcher in 
the Beer and iTunes-Amazon datasets. \rev{We also implemented and ran DeepER~\cite{Ebraheem:2018:DeepER}, 
which is another RNN-based EM solution.}
We denote by DeepMatcher+ (or simply DM+)
the best F1 scores among DeepMatcher and these works aforementioned.
\rev{We summarize in Appendix \ref{sec:breakdown} the implementation details 
and performance of each method.}
\item \textbf{\system: } This is the full version of our system with all 3 optimizations,
domain knowledge (DK), TF-IDF summarization (SU), and data augmentation (DA) turned on. See the details below.
\item \textbf{\system (DA): } This version only turns on the DA (with MixDA)
and SU but does not have the DK optimization.
We apply one of the span-level or attribute-level DA operators listed in Table \ref{tab:da} with
the entry\_swap operator. 
We compare the different combinations and report the best one.
Following \cite{miao2020snippext}, we apply MixDA with the interpolation parameter 
$\lambda$ sampled from a Beta distribution $\text{Beta}(0.8, 0.8)$.
\item \textbf{\system (DK): } With only the DK and SU optimizations on, this version
of \system\ is expected to have lower F1 scores but train much faster.
We apply the span-typing to datasets of each domain according to Table \ref{tab:spans} and 
apply the span-normalization on the number spans.
\item \textbf{Baseline: } This base form of \system\ corresponds
simply to fine-tuning a pre-trained LM on the EM task. We did not apply any optimizations on the baseline.
\rev{
For each ER-Magellan dataset, we tune the LM for the baseline and
found that RoBERTa generally achieves the best performance. Thus, we use
RoBERTa in the other 3 \system\ variants (\system, \system (DA), and \system (DK)) by default
across all datasets. The Company dataset is the only exception, where we found that the BERT model performs the best.
For the WDC benchmark, since the training sets are large,
we use DistilBERT across all settings for faster training. }
% This is the only variant without the %summarization (SU) optimization. 
%\wctan{I don't understand the last sentence} %\yuliang{Baseline is the only variant of %\system\ 
%without the summarization optimization. %\system, \system(DA), and \system(DK) all %have SU.}
% We pick DistilBERT instead of  larger models such as BERT or ALBERT because DistilBERT is faster to train and 
% it also makes a tougher comparison for \system{} since
% larger models are generally perceived to have more powerful language understanding capabilities \cite{yang2019xlnet,liu2019roberta,lan2019albert}.
%\wctan{added last sentence.pls add %references} \yuliang{We have said earlier %that we use DistilBERT 
%because we want to speed up the training %process. This sounds contradicting.}
%\wctan{take a look now} \yuliang{good.}
\end{itemize}
There is a concurrent work~\cite{brunner2020entity}, which also applies pre-trained LM to the entity matching
problem. The proposed method is similar to the baseline method above, but due to the difference in the
evaluation methods (\cite{brunner2020entity} reports the best epoch on the test set, instead of the validation set),
% \yoshi{Do we want to explicitly mention this ([6] reports the best epoch on the test set, instead of the validation set) in the main paper? I think what's more important is that Baseline in Table 5 is their method [6] based on our experimental settings.} \yuliang{I do agree that the test vs. valid set statement looks odd here. My (sneaky) plan is to keep the statement in the revision to emphasize the difference to the reviewers, but we can drop it after the paper is accepted. What do you think?}\yoshi{The plan sounds good to me.} 
the reported results in \cite{brunner2020entity} is not directly comparable.
We summarize in Appendix \ref{sec:concurrent} the difference between \system\ and \cite{brunner2020entity}
and explain why the reported results are different.

\vspace{-1mm}
\subsection{Main results}

\iffalse %%%
Table \ref{tab:magellan} shows the results on the ER-Magellan datasets.
Overall, \system\ achieves significantly higher F1 scores than the SOTA results (DeepMatcher+).
\system\ outperforms DeepMatcher+ in 10/13 cases and by up to 25\% (Dirty, Walmart-Amazon).
On the 3 cases that \system\ performs slightly worse than DeepMatcher+,
as we show in Table \ref{tab:albert}, these small gaps can be filled
by replacing DistilBERT with larger pre-trained LMs such as BERT or ALBERT.
\fi %%%

Table~\ref{tab:magellan} shows the results of the ER-Magellan datasets. Overall, \system{} (with optimizations) achieves significantly higher F1 scores than the SOTA results (DM+). \system{} without optimizations (i.e., the baseline) achieves comparable results with DM+.
\system\ outperforms DM+ in  \rev{all 13} cases and by up to \rev{31\%} (Dirty, Walmart-Amazon) while the baseline outperforms DM+ in \rev{12/13 cases except for the Company dataset with long text}.
% On the 3 cases that \system\ performs slightly worse than DeepMatcher+,
% it turns out that using a larger
% pre-trained LMs such as BERT or ALBERT helps fill the gaps (see Table \ref{tab:albert}). These initial results led us to believe that larger pre-trained language models will further improve \system's results and we leave as future work to further verify this hypothesis.
%in general and we are conducting further %experiments to investigate this hypothesis.
%We are also conducting
%these small gaps can be filled
%by replacing DistilBERT with larger %pre-trained LMs such as BERT or ALBERT.  
%\wctan{pls read} \yuliang{I don't think we %should say ``we are conducting further %experiments...''.
%It sounds like the work is incomplete. I %suggest saying something like 
%``Our initial results showed that larger LMs %improve \system's results further but we leave that as future work'' }
% \wctan{take a look} \yuliang{good.}

In addition, we found that \system\ is better at datasets with small training sets.
Particularly, the average improvement on the 7 smallest datasets is \rev{15.6\%} vs. \rev{1.48\%} on average on the rest of datasets. 
\system\ is also more robust against data noise than DM+. In the 4 dirty datasets,
the performance degradation of \system\ is only \rev{0.57} on average
while the performance of DM+ degrades by 8.21.
% The improvement is 10.06\% on average for the noisy settings and 5.51\% for all 13 cases. 
These two properties make \system\ more attractive in practical EM settings.

\rev{
Moreover, in Appendix \ref{sec:label_efficiency}, we show an evaluation of \system's
label efficiency on 5 of the ER-Magellan medium-size datasets. In 4/5 cases, 
when trained on less than 20\% of the original training data,
\system\ is able to achieve close or even better performance than DM+ when the full
training sets are in use.}

\system\ also achieves promising results on the WDC datasets (Table \ref{tab:wdc}).
\system\ achieves the highest F1 score of 94.08 when using all the 215k training data, outperforming
the previous best result by 3.92. Similar to what we found in the ER-Magellan datasets,
the improvements are higher on settings with fewer training examples (to the right of Table \ref{tab:wdc}).
The results also show that \system\ is more \emph{label efficient} than DeepMatcher.
For example, when using only 1/2 of the data (Large), \system\ already outperforms DeepMatcher with all the training data (xLarge) by 2.89 in All. When using only 1/8 of the data (Medium), the performance is within 1\% close to DeepMatcher's F1 when 1/2 of the data (Large) is in use.
The only exception is the shoes category. This may be caused by the large gap of the positive label ratios 
between the training set and the test set (9.76\% vs. 27.27\% according to Table \ref{tab:wdc-dataset}).

\setlength{\tabcolsep}{2.3pt}
\begin{table}[!ht]
\small
\centering
\caption{\small F1 scores on the ER-Magellan EM datasets. 
The numbers of DeepMatcher+ (DM+) are the highest available found in \protect\cite{Fu:2019:End2End,Kasai:2019:LowResourceER,Mudgal:2018:DeepMatcher} \rev{or re-produced by us}.
% \system(DA) improves Baseline by 0.53 on average thus is not shown in this table.
}\label{tab:magellan}
\vspace{-3mm}
\resizebox{0.47\textwidth}{!}{ 
\begin{tabular}{ccccccc}
\toprule
Datasets           & DM+ & 
\system & 
\begin{tabular}{c}
\system     \\
(DA)
\end{tabular}
 & \begin{tabular}{c}
\system     \\
(DK)
\end{tabular}
&
Baseline & Size    \\ \midrule
Structured         &              &                &            & &          &         \\
Amazon-Google      & 70.7         & 75.58 \green{(+4.88)}       & 75.08  & 74.67      & 74.10    & 11,460  \\
Beer               & 78.8         & 94.37 \green{(+15.57)}       & 87.21  & 90.46      & 84.59    & 450     \\
DBLP-ACM           & 98.45         & 98.99 \green{(+0.54)}       &  99.17 & 99.10      & 98.96    & 12,363  \\
DBLP-Google & 94.7         & 95.6 \green{(+0.9)}    &   95.73   & 95.80      & 95.84    & 28,707  \\
Fodors-Zagats      & 100          & 100.00 \green{(+0.0)}     &  100.00   & 100.00      & 98.14    & 946     \\
iTunes-Amazon      & 91.2         & 97.06 \green{(+5.86)}    &  97.40    & 97.80      & 92.28    & 539     \\
Walmart-Amazon     & 73.6         & 86.76 \green{(+13.16)}    &  85.50    & 83.73      & 85.81    & 10,242  \\ \midrule
Dirty              &              &                &          &  &          &         \\
DBLP-ACM           & 98.1         & 99.03 \green{(+0.93)}     & 98.94    & 99.08      & 98.92    & 12,363  \\
DBLP-Google & 93.8         & 95.75 \green{(+1.95)}  &  95.47      & 95.57      & 95.44    & 28,707  \\
iTunes-Amazon      & 79.4         & 95.65 \green{(+16.25)} &  95.29       & 94.48      & 92.92    & 539     \\
Walmart-Amazon     & 53.8         & 85.69 \green{(+31.89)} &  85.49  & 80.67      & 82.56    & 10,242  \\ \midrule
Textual            &              &                &    &         &          &         \\ 
Abt-Buy            & 62.8         & 89.33 \green{(+26.53)} &  89.79        & 81.69      & 88.85    & 9,575   \\
Company            & 92.7         & 93.85 \green{(+1.15)} &    93.69     & 93.15      & 41.00    & 112,632 \\ \bottomrule
\end{tabular}}
\vspace{-2mm}
\end{table}

% \begin{table}[!ht]
% \vspace{-2mm}
% \small
% \centering
% \caption{\small F1 scores of \system\ with the base BERT and ALBERT models on the 3 datasets where
% \system\ with DistilBERT does not outperform DeepMatcher+ (DM+), the SOTA matching models. }\label{tab:albert}
% \begin{tabular}{cccccc}
% \toprule
% Datasets      & DM+    & Ditto (BERT) & delta & Ditto (ALBERT) & delta \\ \midrule
% DBLP-Google   & 94.70  & 94.80        & \green{(+0.10)}  & 94.73          & \green{(+0.03)}  \\
% Fodors-Zagats & 100.00 & 100.00       & 0.00  & 100.00         & 0.00  \\
% Company       & 92.70  & 93.15        & \green{(+0.45)}  & 92.89          & \green{(+0.19)} \\ \bottomrule
% \end{tabular}
% \vspace{-2mm}
% \end{table}

\setlength{\tabcolsep}{3.4pt}
\begin{table}[!ht]
\centering
\small
\caption{\small F1 scores on the WDC product matching datasets. The numbers for DeepMatcher (DM) are taken
from~\protect\cite{Primpeli:2019:WDC}.}\label{tab:wdc}
\vspace{-2mm}
\begin{tabular}{ccccccccc} \toprule
Size                       & \multicolumn{2}{c}{xLarge (1/1)} & \multicolumn{2}{c}{Large (1/2)} & \multicolumn{2}{c}{Medium (1/8)} & \multicolumn{2}{c}{Small (1/20)} \\
Methods                    & DM             & Ditto         & DM             & Ditto          & DM              & Ditto          & DM              & Ditto          \\ \midrule
\multirow{2}{*}{Computers} & 90.80          & 95.45         & 89.55          & 91.70          & 77.82           & 88.62          & 70.55           & 80.76          \\
                           & \multicolumn{2}{c}{\green{+4.65}}       & \multicolumn{2}{c}{\green{+2.15}}        & \multicolumn{2}{c}{\green{+10.80}}        & \multicolumn{2}{c}{\green{+10.21}}        \\ \midrule
\multirow{2}{*}{Cameras}   & 89.21          & 93.78         & 87.19          & 91.23          & 76.53           & 88.09          & 68.59           & 80.89          \\
                           & \multicolumn{2}{c}{\green{+4.57}}       & \multicolumn{2}{c}{\green{+4.04}}        & \multicolumn{2}{c}{\green{+11.56}}        & \multicolumn{2}{c}{\green{+12.30}}        \\ \midrule
\multirow{2}{*}{Watches}   & 93.45          & 96.53         & 91.28          & 95.69          & 79.31           & 91.12          & 66.32           & 85.12          \\
                           & \multicolumn{2}{c}{\green{+3.08}}       & \multicolumn{2}{c}{\green{+4.41}}        & \multicolumn{2}{c}{\green{+11.81}}        & \multicolumn{2}{c}{\green{+18.80}}        \\ \midrule
\multirow{2}{*}{Shoes}     & 92.61          & 90.11         & 90.39          & 88.07          & 79.48           & 82.66          & 73.86           & 75.89          \\
                           & \multicolumn{2}{c}{\red{-2.50}}      & \multicolumn{2}{c}{\red{-2.32}}       & \multicolumn{2}{c}{\green{+3.18}}         & \multicolumn{2}{c}{\green{+2.03}}         \\ \midrule
\multirow{2}{*}{All}       & 90.16          & 94.08         & 89.24          & 93.05          & 79.94           & 88.61          & 76.34           & 84.36          \\
                           & \multicolumn{2}{c}{\green{+3.92}}       & \multicolumn{2}{c}{\green{+3.81}}        & \multicolumn{2}{c}{\green{+8.67}}         & \multicolumn{2}{c}{\green{+8.02}} \\ \bottomrule   
\end{tabular}
\vspace{-5mm}
\end{table}

\begin{figure*}[]
    \centering
    \includegraphics[width=0.98\textwidth]{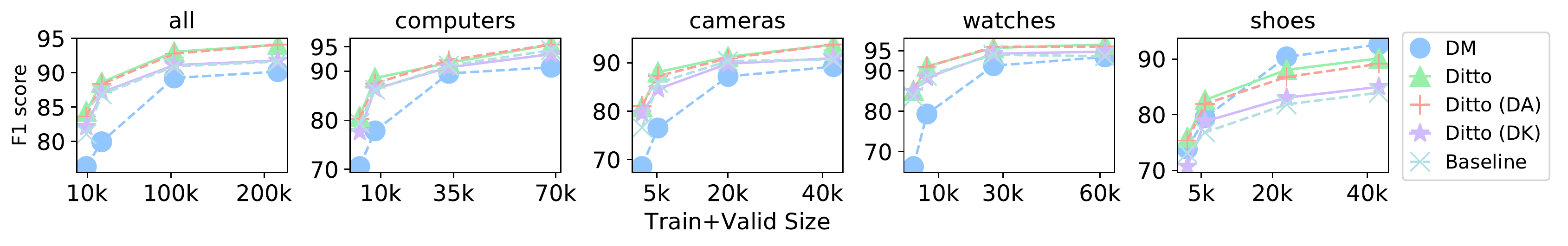}
    \vspace{-3mm}
    \caption{\small F1 scores on the WDC datasets of different versions of \system. \textbf{DM}: DeepMatcher.}
    \label{fig:wdc}
    \vspace{-3mm}
\end{figure*}

\vspace{-1mm}
\subsection{Ablation study}

Next, we analyze the effectiveness of each component (i.e., LM, SU, DK, and DA)
by comparing \system\ with its variants without these optimizations.
The results are shown in Table \ref{tab:magellan} and Figure \ref{fig:wdc}.

The use of a pre-trained LM contributes to a large portion of the performance gain.
In the ER-Magellan datasets (excluding Company), the average improvement of the baseline 
compared to DeepMatcher+ is \rev{7.75}, which accounts for \rev{78.5\%} of the improvement of the full \system\ (\rev{9.87}).
While DeepMatcher+ and the baseline \system\ (essentially fine-tuning DistilBERT) 
are comparable on the Structured datasets, the baseline performs much better on all the Dirty datasets
and the Abt-Buy dataset. This confirms our intuition that the language understanding
capability is a key advantage of \system\ over existing EM solutions.
%\yuliang{Added these two sentences.}
The Company dataset is a special case because the length of the company articles (3,123 words on average)
is much greater than the max sequence length of 256. 
The SU optimization increases the F1 score of this dataset from 41\% to over \rev{93\%}.
In the WDC datasets, across the 20 settings, LM contributes to 3.41 F1 improvement on average, which explains 55.3\% of improvement of the full \system\ (6.16).

The DK optimization is more effective on the ER-Magellan datasets.
Compared to the baseline, the improvement of \system(DK) is \rev{1.08} on average and 
is up to \rev{5.88} on the Beer dataset
while the improvement is only 0.22 on average on the WDC datasets.
We inspected the span-typing output and found that only 66.2\% of entry pairs have spans of the same type. This is caused by the current NER module not extracting product-related spans with the correct types. We expect DK to be more effective if we use
an NER model trained on the product domain.

DA is effective on both datasets and more significantly on the WDC datasets.
The average F1 score of the full \system\ improves upon \system(DK) (without DA) by \rev{1.39} and 2.53
respectively in the two datasets.
In the WDC datasets, we found that the span\_del operator always performs the best
while the best operators are diverse in the ER-Magellan datasets. We list the best operator for each dataset in Table \ref{tab:bestda}. We note that there is a large space
of tuning these operators
(e.g., the MixDA interpolation parameter, maximal span length, etc.)
and new operators to further improve the performance.
% Finding the best DA operators for EM is future work beyond the scope of this paper.

\begin{table}[!htb]
\vspace{-2mm}
\small
\centering
\caption{\small Datasets that each DA operator achieves the best performance.
The suffix (S)/(D) and (Both) denote the clean/dirty version of the dataset or both of them. All operators are applied with the entry\_swap operator.}
 \vspace{-2mm}
\label{tab:bestda}
\begin{tabular}{cc}
\toprule
\textbf{Operator} & \textbf{Datasets} \\ \midrule
span\_shuffle & DBLP-ACM (Both), DBLP-Google (Both), Abt-Buy      \\
span\_del     & Walmart-Amazon(D), Company, all of WDC          \\
attr\_del     & Beer, iTunes-Amazon(S), Walmart-Amazon(S) \\
attr\_shuffle & Fodors-Zagats, iTunes-Amazon(D) \\ \bottomrule
\end{tabular}
\vspace{-2mm}
\end{table}

% without DK

% without DA

% Comparing different DA operators

% \subsection{Error analysis}

% \yuliang{not sure what to write yet.}
\section{Case Study: Employer Matching}\label{sec:casestudy}

We present a case of applying \system\ to a real-world EM task.
%We put \system\ in the perspective of %a complete EM pipeline to demonstrate %how to adopt \system\ and the %immediate performance gain.
An online recruiting platform
would like to join its internal employer records with newly collected public records to enable downstream aggregation tasks.
Given two tables $A$ and $B$ (internal and public) of employer records,
the goal is to find, for each record in table $B$,
a record in table $A$ that represents the same employer.
Both tables have 6 attributes: \textsf{name, addr, city, state, zipcode,} and \textsf{phone}.
Our goal is to find matches with both high precision and recall.

\smallskip
\noindent
\textbf{Basic blocking. } Our first challenge is size of the datasets. 
Table \ref{tab:employer-dataset} shows that both tables
are of nontrivial sizes even after deduplication. 
% Thus, a naive pairwise comparison is not feasible. % because there are in total 49 billion candidate pairs.
The first blocking method we designed is to only match companies with \emph{the same zipcode}. 
However, since 60\% of records in Table $A$ do not have the \textsf{zipcode} attribute
and some large employers have multiple sites,
we use a second blocking method that returns for each record in Table $B$ the 
top-20 most similar records in $A$ ranked by the TF-IDF cosine similarity of \textsf{name} and \textsf{addr} attributes.
We use the union of these two methods as our blocker, which produces 10 million candidate pairs.

\begin{table}[!ht]
\small
\centering
\caption{\small Sizes of the two employer datasets to be matched.}\label{tab:employer-dataset}
 \vspace{-4mm}
\begin{tabular}{cccccc} \toprule
       & \multicolumn{2}{c}{TableA} & \multicolumn{2}{c}{TableB} & \#Candidates \\
       & original   & deduplicated  & original   & deduplicated  & Basic blocking \\ \midrule
Size & 789,409    & 788,094       & 412,418    & 62,511     & 10,652,249\\ \bottomrule
\end{tabular}
    \vspace{-2mm}
\end{table}

\smallskip
\noindent
\textbf{Data labeling. } We labeled 10,000 pairs sampled 
from the results of each blocking method (20,000 labels in total).
We sampled pairs of high similarity with higher probability
to increase the difficulty of the dataset to train more robust models.
The positive rate of all the labeled pairs is 39\%.
We split the labeled pairs into training, validation, and test sets by the ratio of \textsf{3:1:1}.
%\wctan{what is the proportion of positive and negative pairs in the 20K?}

\smallskip
\noindent
\textbf{Applying \system. } The user of \system\ does not need to extensively tune the hyperparameters
but only needs to specify the domain knowledge and choose a data augmentation operator.
We observe that the street number and the phone number are both useful signals for matching.
Thus, we implemented a simple $\recognizer$ that tags the first number string in the \textsf{addr}
attribute and the last 4 digits of the \textsf{phone} attribute. 
Since we would like the trained model to be robust against the large number of 
missing values, we choose the attr\_del operator for data augmentation.

\begin{figure}[!ht]
    \vspace{-2mm}
    \centering
    \includegraphics[width=0.46\textwidth]{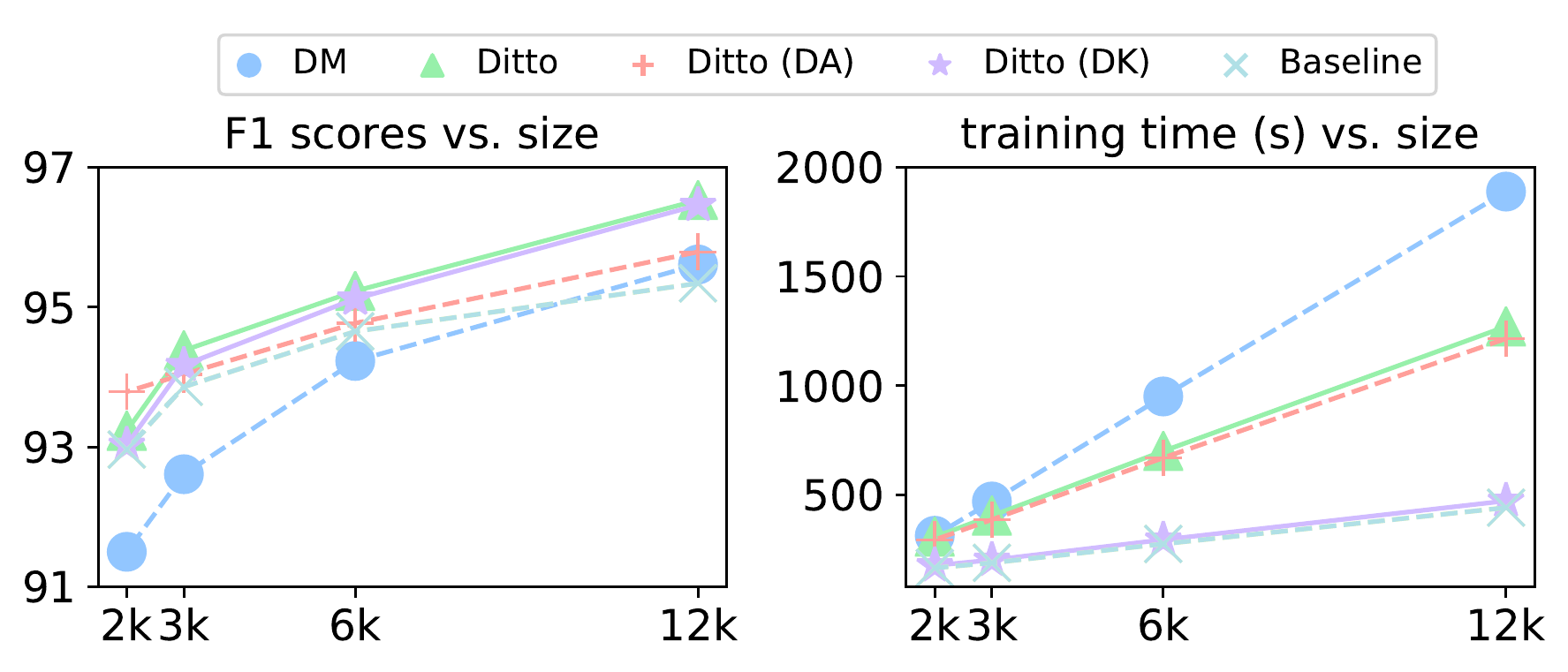}
    \vspace{-4mm}
    \caption{\small F1 and training time for the employer matching models.}
    \label{fig:employer}
    \vspace{-3mm}
\end{figure}

We plot the model's performance in Figure \ref{fig:employer}. \system\ achieves
the highest F1 score of 96.53 when using all the training data.
\system\ outperforms DeepMatcher (DM) in F1 and trains faster (even when using MixDA) than DeepMatcher across different 
training set sizes.

\smallskip
\noindent
\textbf{Advanced blocking. } Optionally, before applying the trained model to 
all the candidate pairs, we can use the labeled data to improve the basic blocking method.
We leverage Sentence-BERT~\cite{reimers2019sentence}, a variant of the BERT model
that trains sentence embeddings for sentence similarity search.
The trained model generates a high-dimensional (e.g., 768 for BERT) vector 
for each record.
Although this model has a relatively low F1 (only 92\%) 
thus cannot replace \system,
we can use it with vector similarity search
to quickly find record pairs that are likely to match.
We can greatly reduce the matching time by only testing those pairs of high cosine similarity.
% {\color{red} A similar idea was used in \cite{Ebraheem:2018:DeepER}.}\yoshi{I think we can remove this sentence as we mention that we're inspired by DeepER in the Related Work section.} %\yuliang{moved this sentence to related work?}
We list the running time for each module in Table \ref{tab:time}.
With this technique, the overall EM process is accelerated by 3.8x
(1.69 hours vs. 6.49 hours with/without advanced blocking).

\begin{table}[!ht]
%    \vspace{-3mm}
\small
\centering
\caption{\small Running time for blocking and matching with \system. 
Advanced blocking consists of two steps: computing the representation of each record with 
Sentence-BERT~\protect\cite{reimers2019sentence} (Encoding)
and similarity search by blocked matrix multiplication~\protect\cite{abuzaid2019index} (Search).
With advanced blocking, we only match each record
with the top-10 most similar records according to the model.} \label{tab:time}
 \vspace{-2mm}
\begin{tabular}{cccccc} \toprule
 & Basic & Encoding & Search  & \multicolumn{2}{c}{Matching}            \\
 & Blocking & (GPU) & (CPU) &  (top-10) & (ALL)            \\ \midrule
Time (s) & 537.26  & 2,229.26   & 1,981.97       & 1,339.36    & 22,823.43 \\ \bottomrule
\end{tabular}
    \vspace{-2mm}
\end{table}
\section{Related Work \rev{and Discussion}}\label{sec:related}

%Entity Matching (EM), also known as entity resolution, duplication detection %\jinfeng{too broad} 
%\yuliang{I saw this in paper's title, like \cite{blocking2} and two more in the .bib file.}\jinfeng{or Record deduplication as in \cite{blocking2}?}
%or entity/record linkage,
%has been a long-standing problem in data
EM solutions
have tackled the blocking problem~\cite{blocking1,blocking2,blocking3,Papadakis:2019:BlockingSurvey,blocking4} and
the matching problem with rules~\cite{dalvi2013optimal,elmagarmid2014nadeef,singh2017synthesizing,wang2011entity}, crowdsourcing~\cite{gokhale2014corleone,karger2011human,wang2012crowder}, or
machine learning~\cite{sarawagi2002interactive,cohen2002learning,bilenko2003adaptive,gokhale2014corleone,Konda:2016:Magellan}. 
%The classical machine learning techniques %used in EM include
%decision trees~\cite{gokhale2014corleone},
%SVM~\cite{bilenko2003adaptive},
%%clustering~\cite{cohen2002learning}, and
%active learning~\cite{cohen2002learning,sar%awagi2002interactive}. 

%~\ci
%integration~\cite{doan2012principles, %doan2005semantic,naumann2010introduction}.
%EM is typically done in two phases: \emph{blocking} and \emph{matching}~\cite{Konda:2016:Magellan}. %\jinfeng{change order? "blocking and matching"}. \yuliang{good catch}  
%Blocking~\cite{blocking1,blocking2,blocking3,Papadakis:2019:BlockingSurvey,blocking4} speeds up EM by retaining only the plausible
%candidate pairs. 

%Like these solutions, the primary focus of \system{} is on the matching phase and uses deep learning methods for this purpose.
%focuses primarily on the matching phase which 
%checks whether each pair is a real match. 
Recently, EM solutions used
deep learning and
achieved promising results~\cite{Ebraheem:2018:DeepER,Fu:2019:End2End,Kasai:2019:LowResourceER,Mudgal:2018:DeepMatcher,Zhao:2019:AutoEM}.
DeepER~\cite{Ebraheem:2018:DeepER} trains EM models based on the LSTM~\cite{lstm} neural network architecture with word embeddings such as GloVe~\cite{glove}.
% DeepER also proposed a blocking technique that represents each entry by the the LSTM's output vector
% and uses Locality Sensity Hashing~\cite{datar2004locality} for similarity search.
DeepER also proposed a blocking technique to represent each entry by the LSTM's output.
Our advanced 
blocking technique based on Sentence-BERT~\cite{reimers2019sentence}, 
described in Section \ref{sec:casestudy}, is inspired by this. 
Auto-EM~\cite{Zhao:2019:AutoEM} improves deep learning-based EM models by pre-training the EM model
on an auxiliary task of entity type detection. \system\ also leverages transfer learning
by fine-tuning pre-trained LMs, which are more powerful in language understanding. We did not compare \system\ with Auto-EM in experiments
because the entity types required by Auto-EM are not available in our benchmarks.
%\wctan{I don't understand the prev sentence.}
%\yuliang{AutoEM requires each entity has a type like computers, shoes, persons etc.
%I think this is useful only when the dataset has mixed types like in the WDC datasets.
%The Magellan-ER dataset all entries are of the same type, so adding this pre-training is not going to help.}
However,
we expect that pre-training \system\ with EM-specific data/tasks can improve the performance of \system{} further
and is part of our future work. 
%\yoshi{Sentence-BERT or SBERT? Make it consistent with the name in Section 6.}
DeepMatcher introduced a design 
space for applying deep learning to EM.
%, by including 
%soft-alignment~\cite{bahdanau2014neural}, %attention mechanism~\cite{bahdanau2014neura%l},
%hybrid models~\cite{wang2016compare}, and %variants of RNN~\cite{rnn}.
%It also uses FastText~\cite{fasttext} to %better capture sub-word information and
%achieved SOTA results on a large number of %EM datasets. %\yuliang{added this.}
%\system{} focuses primarily on the matching phase using pre-trained Transformers-based LMs. 
Following their template architecture, one can think of \system{} as replacing
both the attribute embedding and similarity representation components in the architecture with a single pre-trained LM such as BERT, thus providing a much simpler overall architecture. 
% stopped here.
%input difference
%BERT difference
%One difference between \system{} and systems %such as DeepER and DeepMatcher is the way data %entries are ingested.

All systems, Auto-EM, DeepER, DeepMatcher, and \system{} formulate
matching as a binary classification problem.
The first three take a pair of data entries of the same arity as input and aligns the attributes before passing them to 
the system for matching. On the other hand, \system{} serializes both data entries as one input with structural tags intact. This way, data entries of different schemas can be uniformly ingested,  including hierarchically formatted data such as those in JSON. %The single input serialization scheme %also facilitates optimizations where %we inject domain knowledge by tagging %values of different granularity in %the input and applying data %augmentation. 
Our serialization scheme is not only applicable to \system, but also to other systems such as DeepMatcher. 
In fact, we serialized data entries to DeepMatcher under one attribute using our scheme and observed that DeepMatcher improved by as much as 5.2\% on some datasets.
%\yuliang{the last statement is a bit dangerous because DK %and DA can be applied to AutoEM, DeepER, and DeepMatcher as %well.}

\rev{
A concurrent work~\cite{brunner2020entity} also applies pre-trained LMs to the entity matching problem
and achieves good performance. While the proposed method in \cite{brunner2020entity} is similar to the baseline
version of \system, \system\ can be further optimized using domain knowledge, data augmentation, and summarization.
We also present a comprehensive experiment analysis on more EM benchmarks using a more standard evaluation method.
We provide a detailed comparison between \system\ and \cite{brunner2020entity}
in Appendix \ref{sec:concurrent}.
}

%%% say something about easy to do optimizations
\iffalse %%%%
Like DeepER and DeepMatcher, \system\ formulates matching as a binary classification problem to apply deep learning techniques and in \system's case,
Transformers-based models~\cite{vaswani2017attention} which have 
more sophisticated attention mechanisms enabling better alignment between
data entry pairs are applied. 
A model architecture of good potential is DeepMatcher with 
the RNN+FastText components replaced with a pre-trained LM such as BERT, which
we leave as future work. \yuliang{added the sentence.}
\fi %%%

% Transformers-based language model is a recent hot topics in NLP.
% \yuliang{TODO}
% To the best of our knowledge, \system\ is the first system that applies
% pre-trained LMs to EM.

%We have reviewed in Section %\ref{sec:architecture} the pre-trained LMs
%that \system\ fine-tunes to train the matching %model.
%\system\ optimizes the performance of the LM %further by using 
%domain knowledge and data augmentation. % \jinfeng{Can DK and DA improve DeepMatcher and DeepER?}. 
% External knowledge is known to be effective in improving neural network models in NLP tasks~\cite{chen2017neural,sun2019ernie}.
% Instead of directly modifying the network architecture~\cite{Wang:2019:KGAT,Yang:2017:KBLSTM} or the loss function~\cite{zhang-etal-2019-ernie} 
% to incorporate domain knowledge, \system\ modularizes the way domain knowledge is
External knowledge is known to be effective in improving neural network models in NLP tasks~\cite{Yang:2017:KBLSTM,chen2017neural,sun2019ernie,Wang:2019:KGAT}. 
To incorporate domain knowledge, \system\ modularizes the way domain knowledge is
incorporated by allowing users to specify and customize rules for preprocessing input entries.
Data augmentation (DA) has been extensively studied in computer vision and has recently received more attention in NLP~\cite{miao2020snippext,wei2019eda,xie2019unsupervised}.
We designed a set of DA operators suitable for EM and
apply them with MixDA~\cite{miao2020snippext}, a recently proposed DA strategy based on
convex interpolation. To our knowledge, this is the first time data augmentation has been applied to EM.

\rev{
Active learning is a recent trend in EM to train high-quality 
matching models with limited labeling resources~\cite{gurajada2019learning,Kasai:2019:LowResourceER,meduri2020comprehensive,qian2017active}.
Under the active learning framework, the developer interactively labels a small set of examples to improve the model while the updated model is used to sample new examples for the next labeling step. Although active learning's goal of improving label efficiency aligns with data augmentation in \system, they are different solutions, which can be used together; active learning requires human interaction in each iteration, whereas data augmentation does not.
According to \cite{meduri2020comprehensive}, one needs to adjust the model size and/or the training process such that the response time becomes acceptable for user interaction in active learning. Thus, we consider applying it to \system\ is not straightforward because of the relatively long fine-tuning time of the \system.
We leave this aspect to future development of \system.}
% \yoshi{I directly edited this paragraph and this may become too long. My point is to tell the reviewer that DA and Active Learning are different (and orthogonal) solutions. Please check and feel free to revert.} 
% \yuliang{looks good.}

\smallskip
\noindent
\rev{\textbf{Discussion. } Like other deep learning-based EM solutions,
\system\ requires a non-trivial amount of labeled training examples 
(e.g., the case study requires 6k examples to achieve 95\% F1) and \system's DA and DK optimizations help reduce
%. 
%The DA and DK optimizations aim at reducing 
the labeling requirement to some extent.
Currently, the LMs that we have tested in \system\ 
%\yoshi{the LMs that Ditto/we use? the LMs that we test? If Ditto (theoretically) can support other pretrained LMs pretrained on domain-specific and/or non-English resources, the word "support" may not describe our intent correctly.} \yuliang{good point. How about now?}
are pre-trained on general English text corpora and 
thus might not capture well EM tasks with a lot of numeric data and/or
specific domains such as the scientific domain. For domain-specific tasks,
a potential solution is to leverage specialized LMs such as SciBERT~\cite{beltagy2019scibert} 
or BioBERT~\cite{lee2020biobert} trained on scientific and biology corpus respectively.
For numeric data, a good candidate solution would be a hybrid neural network similar to \cite{herzig2020tapas,yin2020tabert} that combines the numeric features
with the textual features.}

\section{Conclusion}\label{sec:conclusion}

We present \system, \rev{an EM system} 
based on fine-tuning pre-trained Transformer-based language models.
\system\ uses a simple architecture to leverage pre-trained LMs
and is further 
% \jinfeng{So AutoEM does not belong to pre-training models?} \yuliang{it is pre-trained but I think it is not a ``language'' model.}. 
optimized by injecting domain knowledge, text summarization, and data augmentation.
Our results show that it outperforms existing EM solutions on all three benchmark datasets with significantly less training data.
\system's good performance can be attributed to the improved language understanding capability mainly through pre-trained LMs, the more accurate text alignment guided by the injected knowledge, and the data invariance properties
learned from the augmented data. 
We plan to further explore our design choices for injecting domain knowledge, text summarization, and data augmentation. 
%We plan to investigate other forms of domain knowledge that can be added and composing multiple DA operators in future.
%These results, however, shed light on the future directions of improving EM such as injecting domain knowledge in different forms and composing multiple DA operators. 
In addition, we plan to extend \system\ to other data integration tasks beyond EM, 
such as entity type detection and schema matching with the ultimate goal of building a BERT-like model for tables.
%Our ultimate goal is to develop a model that, 
%by pre-training on a large collection of %structured data, understands the semantic %meaning of data and can be used in a wide range %of downstream data management tasks.

%\newpage
%\balance
\bibliographystyle{ACM-Reference-Format}
\bibliography{paper}

\newpage

\appendix 

\section{Architecture of the Pre-trained Language models}\label{sec:lm}

\begin{figure}[!ht]
    \centering
    \includegraphics[width=0.49\textwidth]{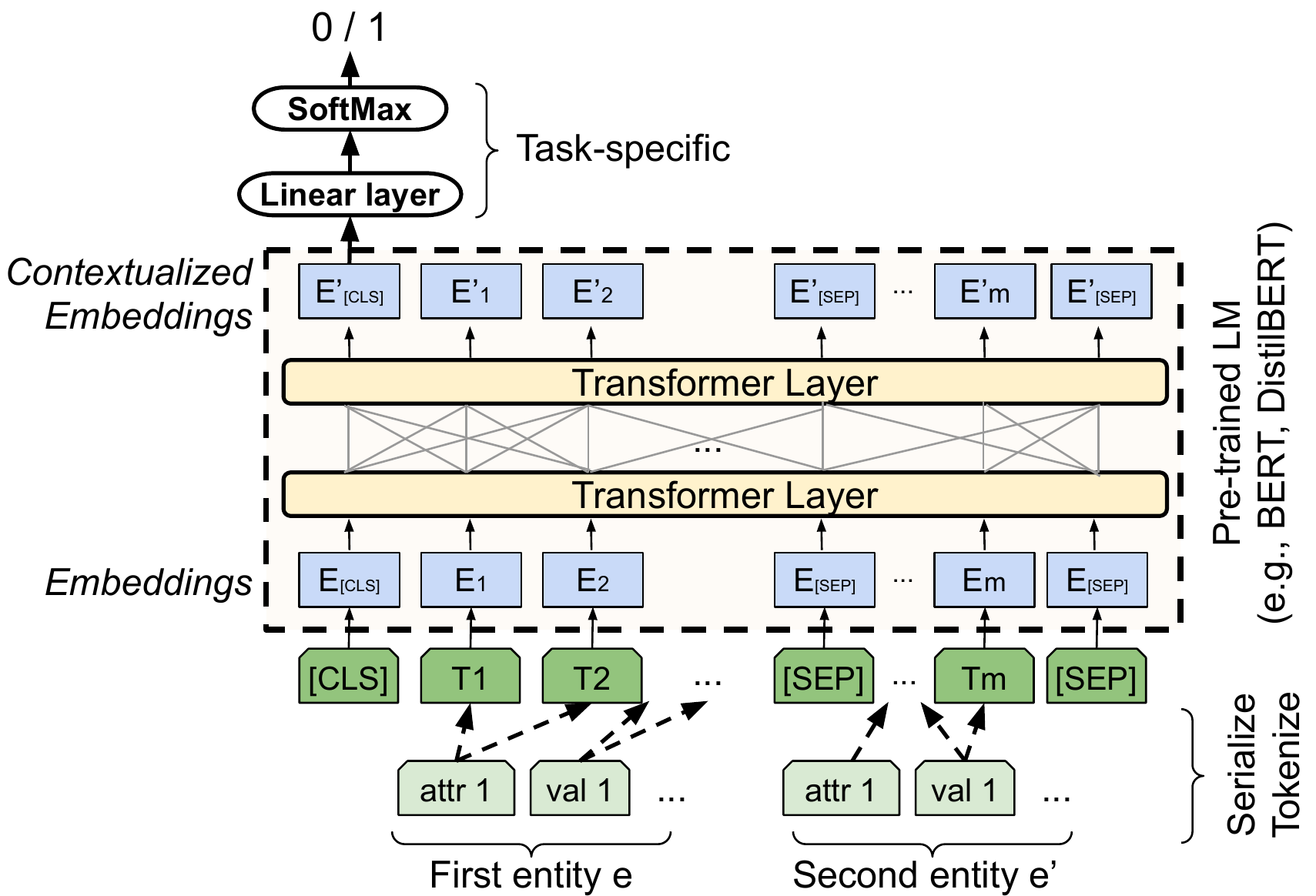}
    \caption{\small \system's model architecture. }
    \label{fig:lm}
\end{figure}

Figure \ref{fig:lm} shows the model architecture of \system's language models
such as BERT~\cite{Devlin:2019:BERT}, DistilBERT~\cite{sanh2019distilbert}, and RoBERTa~\cite{liu2019roberta}.
\system\ serializes the two input entries entries as one sequence
and feeds it to the model as input. The model consists of (1) token embeddings and 
Transformer layers~\cite{wolf2019transformers} from a pre-trained language model (e.g., BERT) and 
(2) task-specific layers (linear followed by softmax). Conceptually, the \textsf{[CLS]} token 
``summarizes'' all the contextual information needed for matching as a contextualized embedding 
vector $E'_{\mathsf{[CLS]}}$ which the task-specific layers take as input for classification.
    
\begin{table*}[!ht]
\small
\centering
\caption{Baseline results from different sources.}\label{tab:breakdown}
\begin{tabular}{cccccccc} \toprule
                   & \begin{tabular}[c]{@{}c@{}}DeepER \\ (reproduced)\end{tabular} & \begin{tabular}[c]{@{}c@{}}DM \\ (reproduced)\end{tabular} & \begin{tabular}[c]{@{}c@{}}DM \\ (reported in \cite{Mudgal:2018:DeepMatcher})\end{tabular} & \begin{tabular}[c]{@{}c@{}}DM (using \\ Ditto's input)\end{tabular} & \begin{tabular}[c]{@{}c@{}}Magellan \\ (reported in \cite{Mudgal:2018:DeepMatcher})\end{tabular} & ACL '19 \cite{Kasai:2019:LowResourceER} & IJCAI '19 \cite{Fu:2019:End2End} \\ \midrule
Structured         &                                                                &                                                            &                                                          &                                                                     &                                                                &         &           \\
Amazon-Google      & 56.08                                                        & 67.53                                                      & 69.3                                                     & 65.78                                                             & 49.1                                                           & -       & 70.7      \\
Beer               & 50                                                             & 69.23                                                      & 72.7                                                     & -                                                                   & 78.8                                                           & -       & -         \\
DBLP-ACM           & 97.63                                                        & 98.42                                                      & 98.4                                                     & 98.86                                                               & 98.4                                                           & 98.45   & -         \\
DBLP-Scholar & 90.82                                                        & 94.32                                                      & 94.7                                                     & 94.56                                                               & 92.3                                                           & 92.94   & -         \\
Fodors-Zagats      & 97.67                                                        & -                                                          & 100                                                      & -                                                                   & 100                                                            & -       & -         \\
iTunes-Amazon      & 72.46                                                        & 86.79                                                      & 88                                                       & 88                                                                  & 91.2                                                           & -       & -         \\
Walmart-Amazon     & 50.62                                                        & 63.33                                                      & 66.9                                                     & 61.67                                                               & 71.9                                                           & -       & 73.6      \\ \midrule
Dirty              &                                                                &                                                            &                                                          &                                                                     &                                                                &         &           \\
DBLP-ACM           & 89.62                                                        & 97.53                                                      & 98.1                                                     & 96.03                                                               & 91.9                                                           & -       & -         \\
DBLP-Scholar & 86.07                                                        & 92.8                                                       & 93.8                                                     & 93.75                                                             & 82.5                                                           & -       & -         \\
iTunes-Amazon      & 67.80                                                        & 73.08                                                      & 79.4                                                     & 70.83                                                             & 46.8                                                           & -       & -         \\
Walmart-Amazon     & 36.44                                                        & 47.81                                                      & 53.8                                                     & 48.45                                                             & 37.4                                                           & -       & -         \\ \midrule
Textual            &                                                                &                                                            &                                                          &                                                                     &                                                                &         &           \\
Abt-Buy            & 42.99                                                        & 66.05                                                      & 62.8                                                     & 67.99                                                               & 43.6                                                           & -       & -         \\
Company            & 62.17                                                        & -                                                          & 92.7                                                     & 90.70                                                             & 79.8                                                           & -       & -    
\\ \bottomrule
\end{tabular}
\end{table*}

\section{Training time and prediction time experiments} \label{sec:time}

\begin{figure}[!ht]
%    \vspace{-2mm}
    \centering
    \includegraphics[width=0.48\textwidth]{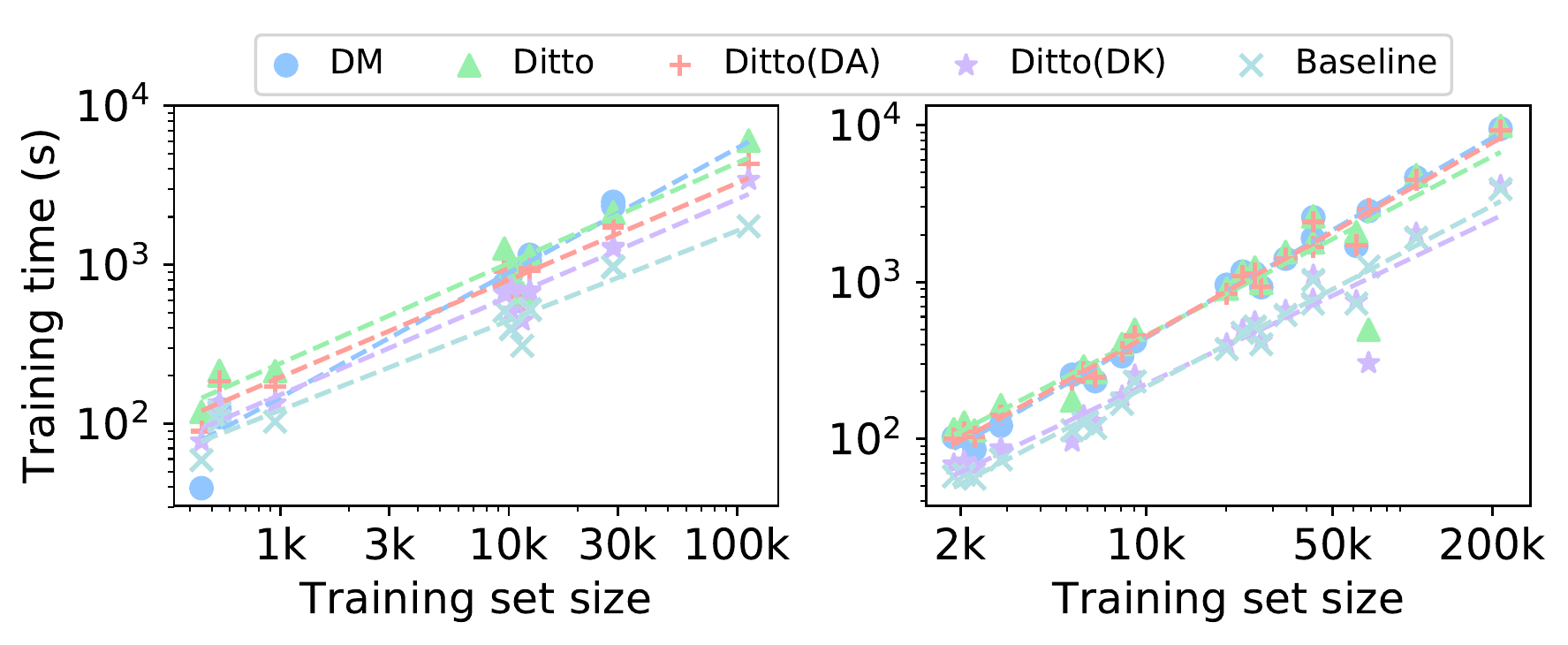}
    \caption{\small Training time vs. dataset size for the ER-Megallan datasets (left) and the WDC datasets (right). Each point corresponds to the training time needed for a dataset using different methods. 
    \rev{\system(DK) and Baseline do not use MixDA thus is faster than the full \system. 
    The DK optimization only adds a small overhead (5\%) to the training time.}
    DeepMatcher (DM) ran out of memory on the Company dataset so the data point is not reported.}
    \label{fig:train_time}
%    \vspace{-2mm}
\end{figure}

We plot the training time
required by DeepMatcher and \system\ in Figure \ref{fig:train_time}.
\rev{The running time for \system\ ranges from 119 seconds (450 examples) to 1.7 hours (113k examples)}.
\system\ has a similar training time to DeepMatcher although 
\rev{the Transformer-based models used by \system are deeper and more complex.}
%\wctan{you mean DistilBERT is more complex? Is this arguable?}
%\yuliang{I think DistilBERT is deeper and Transformers is more complex than RNN.}
The speed-up is due to the \rev{fp16 optimization which is not used by DeepMatcher}. 
\system\ with MixDA is about 2-3x slower than 
\system(DK) without MixDA. This is because MixDA requires additional time for
generating the augmented pairs and computing with the LM twice. 
However, this overhead only affects offline training and does not affect online prediction.

Table \ref{tab:prediction_time} shows \system's average prediction time per entry pair
in each benchmark. The results show that DeepMatcher and \system\ have comparable prediction time.
Also, the DK optimization only adds a small overhead to the
prediction time (less than 2\%). 
The prediction time between the two benchmarks are different because of the difference
in their sequence length distributions.

\begin{table}[!ht]
\small
    \centering
    \caption{\small The average prediction time (ms) per data entry pair of \system. } \label{tab:prediction_time}
\begin{tabular}{cccccc}\toprule
            & \multicolumn{2}{c}{Ditto-DistilBERT} & \multicolumn{2}{c}{Ditto-RoBERTa} & \multirow{2}{*}{DM} \\
            & w. DK                & w/o. DK              & w. DK               & w/o. DK            &                     \\ \midrule
ER-Magellan & 8.01              & 7.87               & 6.82             & 6.78             & 6.62                \\
WDC         & 1.82              & 1.80               & 2.11             & 2.11             & 2.30       \\ \bottomrule
\end{tabular}
\end{table}

\section{Breakdown of the DM+ results and experiments} \label{sec:breakdown}

In this section, we provide a detailed summary of how we obtain the \textbf{DeepMatcher+} (DM+) baseline results.
Recall from Section \ref{sec:setup} that DM+ is obtained by taking the best performance (highest F1 scores)
of multiple baseline methods including
DeepER~\cite{Ebraheem:2018:DeepER}, Magellan~\cite{Konda:2016:Magellan},
DeepMatcher~\cite{Mudgal:2018:DeepMatcher}, and DeepMatcher's follow-up work \cite{Fu:2019:End2End} and \cite{Kasai:2019:LowResourceER}. 

We summarize these baseline results in Table \ref{tab:breakdown} on the ER-Magellan benchmarks 
and explain each method next.

\begin{figure*}[!ht]
    \centering
    \includegraphics[width=1.0\textwidth]{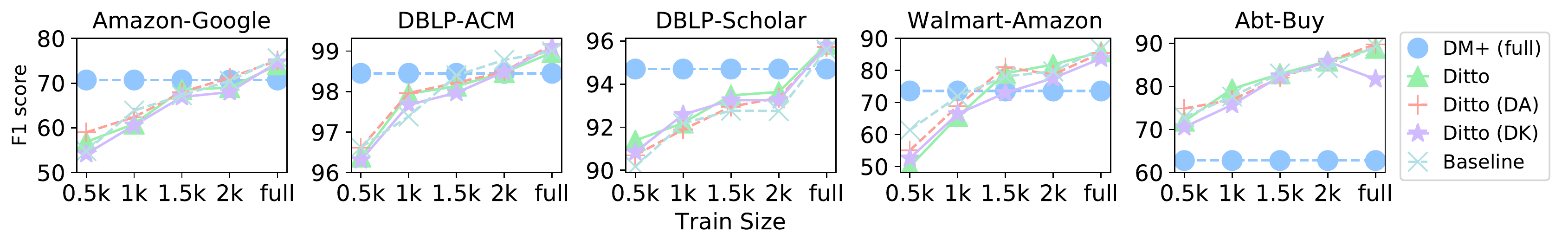}
    % \vspace{-5mm}
    \caption{\small F1 scores on 5 ER-Magellan datasets using different variants of \system. 
    We also plot the score of DeepMatcher+ on the full datasets (denoted as \textbf{DM+(full)}) as reference.
    Recall that full = $\{11460, 12363, 28707, 10242, 9575\}$ for the 5 datasets respectively.}
    % \vspace{-5mm}
    \label{fig:magellan}
\end{figure*}

\smallskip
\noindent
\textbf{DeepER: } The original paper~\cite{Ebraheem:2018:DeepER} proposes a DL-based framework for EM.
Similar to DeepMatcher, DeepER first aggregates both data entries into their vector representations and uses
a feedforward neural network to perform the binary classification based on the similarity of the two vectors.
Each vector representation is obtained either by a simple 
averaging over the GloVe~\cite{glove} embeddings per attribute
or a RNN module over the serialized data entry. DeepER computes the similarity as 
the cosine similarity of the two vectors.
Although \cite{Ebraheem:2018:DeepER} reported results on the Walmart-Amazon, Amazon-Google,
DBLP-ACM, DBLP-Scholar, and the Fodors-Zagat datasets, the numbers are not directly comparable to 
the presented results of \system\ 
because their evaluation and data preparation methods are different 
(e.g., they used k-fold cross-validation while we use the train/valid/test 
splits according to \cite{Mudgal:2018:DeepMatcher}). In our experiments, we implemented DeepER
with LSTM as the RNN module and GloVe for the tokens embeddings as described in \cite{Ebraheem:2018:DeepER}
and with the same hyper-parameters (a learning rate of 0.01 and the Adam optimizer~\cite{kingma2014adam}).
We then evaluate DeepER in our evaluation settings. For each dataset, we report the best results obtained
by the simple aggregation and the RNN-based method.

\smallskip
\noindent
\textbf{DeepMatcher (DM): } We have summarized DM in Section \ref{sec:setup}. In addition to 
simply taking the numbers from the original paper~\cite{Mudgal:2018:DeepMatcher}, we also ran
their open-source version (DM (reproduced)) with the default settings (the Hybrid model with a batch size
of 32 and 15 epochs). The reproduced results are in general lower than the original reported numbers in 
\cite{Mudgal:2018:DeepMatcher} (the 3rd column) because we did not try the other model variants 
and hyperparameters as in the original experiments. The code failed in the Fodors-Zagat and 
the Company datasets because of out-of-memory errors.

In addition, one key difference between
DM and \system\ is that \system\ serializes the data entries while DM does not. One might wonder
if DM can obtain better results by simply replacing its input with the serialized entries produced by \system.
We found that the results do not significantly improved overall, but it is up to 5.2\% in the
Abt-Buy dataset.

\smallskip
\noindent
\textbf{Others: } We obtained the results for Magellan by taking the reported results from \cite{Mudgal:2018:DeepMatcher} and the two follow-up works \cite{Kasai:2019:LowResourceER,Fu:2019:End2End}
of DeepMatcher (denoted as ACL '19 and IJCAI '19 in Table \ref{tab:breakdown}).
We did not repeat the experiments since they have the same evaluation settings as ours.

\section{Label efficiency experiments on the ER-Magellan benchmark} \label{sec:label_efficiency}

We also evaluate the label efficiency of \system\ on the ER-Magellan benchmark.
We conducted the experiments on 5 representative datasets 
(Amazon-Google, DBLP-ACM, DBLP-Scholar, Walmart-Amazon, and Abt-Buy)
of size $\sim$10k to $\sim$30k.
For each dataset, we vary the training set size from 500 to 2,000 and
uniformly sample from the original training set. We then follow the same setting 
as in Section \ref{sec:experiments} to evaluate the 4 variants of \system: 
baseline, \system(DA), \system(DK), and \system.
We summarize the results in Figure \ref{fig:magellan}.
We also plot the result of DM+ trained on the full datasets (denoted as DM+ (full)) as a reference.
As shown in Figure \ref{fig:magellan}, \system\ is able to reach similar or better performance to DM+
on 3 of the datasets (Amazon-Google, DBLP-ACM, and Walmart-Amazon)
with 2,000 train examples (so $\leq20\%$). With only 500 examples, \system\ is able to outperform
DM+ trained on the full data in the Abt-Buy dataset.
These results confirm that \system\ is more label efficient than existing EM solutions.

\section{The difference between Ditto and a concurrent work } \label{sec:concurrent}

% \yuliang{consider moving this section to the related work}

There is a concurrent work \cite{brunner2020entity} which also applies pre-trained LMs
to entity matching and obtained good results. The method proposed in \cite{brunner2020entity}
is essentially identical to the baseline version of \system\ which only serializes 
the data entries into text sequences and fine-tunes the LM on the binary sequence-pair classification task.
On top of that, \system\ also applies 3 optimizations of injecting domain knowledge, data augmentation,
and summarization to further improve the model's performance.
We also evaluate \system\ more comprehensively as we tested \system\ on all the 13 ER-Magellan datasets,
the WDC product benchmark, and a company matching dataset while \cite{brunner2020entity}
experimented in 5/13 of the ER-Magellan datasets.

On these 5 evaluated datasets, one might notice that the reported F1 scores 
in \cite{brunner2020entity} are slightly higher compared to the baseline's F1 scores 
shown in Table \ref{tab:magellan}. The reason is that according to \cite{brunner2020entity},
for each run on each dataset,
the F1 score is computed as the model's \emph{best F1 scores on the test set 
among all the training epochs}, 
% we confirmed this by checking their open-sourced implementation
while we report \emph{the test F1 score of the epoch with the best F1 on the validation set}.
Our evaluation method is more standard since it prevents overfitting the test set 
(See Chapter 4.6.5 of \cite{mitchell1997machine}) 
and is also used by DeepMatcher and Magellan~\cite{Mudgal:2018:DeepMatcher}.
It is not difficult to see that over the same set of model snapshots, the F1 score computed by the 
\cite{brunner2020entity}'s evaluation method would be greater or equal to the F1 score computed
using our method, which explains the differences in the reported values between us and \cite{brunner2020entity}.

Table \ref{tab:lm} summarizes the detailed comparison of the baseline \system, 
the proposed method in \cite{brunner2020entity},
and the full \system. Recall that we construct the baseline by taking the best performing pre-trained model 
among DistilBERT~\cite{sanh2019distilbert}, BERT~\cite{Devlin:2019:BERT}, 
XLNet~\cite{yang2019xlnet}, and \\RoBERTa~\cite{liu2019roberta}
following \cite{brunner2020entity}. Although the baseline \system\ does not outperform \cite{brunner2020entity}
because of the different evaluation method, the optimized \system\ is able to outperform \cite{brunner2020entity}
in 4/5 of the evaluated datasets.

\setlength{\tabcolsep}{2.3pt}
\begin{table}[!ht]
\centering
\small
\caption{\small The F1 scores of the baseline method with different pre-trained LMs.
The first 4 columns are performance of the baseline \system\ using the 4 different LMs. 
We highlight the LM of the best performance on each dataset, 
which form the baseline column in Table \ref{tab:magellan}. 
We turned on the summarization (SU) optimization for the Company dataset
to get F1 scores closer to the full \system. } \label{tab:lm}
\resizebox{0.49\textwidth}{!}{ 
\begin{tabular}{ccccccc}
\toprule
                   & DistilBERT & XLNet & RoBERTa & BERT  & \begin{tabular}{c}
                        Reported  \\
                        in \cite{brunner2020entity} \end{tabular} & Ditto \\ \midrule
Structured         &            &       &         &       &                                                       &       \\
Amazon-Google      & 71.38      & \textbf{74.10} & 65.92   & 71.66 & -                                                     & 75.58     \\
Beer               & 82.48      & 48.91 & 74.23   & \textbf{84.59} & -                                                     & 94.37    \\
DBLP-ACM           & 98.49      & 98.85 & 98.87   & \textbf{98.96} & -                                                     & 98.99     \\
DBLP-Scholar & 94.92      & \textbf{95.84} & 95.46   & 94.93 & -                                                     &  95.6     \\
Fodors-Zagats      & 97.27      & 95.30 & \textbf{98.14}   & 95.98 & -                                                     & 100.0     \\
iTunes-Amazon      & 91.49      & 74.81 & 92.05   & \textbf{92.28} & -                                                     & 97.06     \\
Walmart-Amazon     & 79.81      & 77.98 & \textbf{85.81}   & 81.27 & -                                                     & 86.76     \\ \midrule
Dirty              &            &       &         &       &                                                       &       \\ 
DBLP-ACM           & 98.60      & \textbf{98.92} & 98.79   & 98.81 & 98.90                                                 & 99.03 \\
DBLP-Scholar & 94.76      & 95.26 & \textbf{95.44}   & 94.72 & 95.60                                                 & 95.75 \\
iTunes-Amazon      & 90.12      & 92.70 & \textbf{92.92}   & 92.25 & 94.20                                                 & 95.65 \\
Walmart-Amazon     & 77.91      & 61.73 & \textbf{82.56}   & 81.55 & 85.50                                                 & 85.69 \\ \midrule
Textual            &            &       &         &       &                                                       &       \\
Abt-Buy            & 82.47      & 53.55 & \textbf{88.85}   & 84.21 & 90.90                                                 & 89.33 \\
Company            & 93.16      & 71.93 & 85.89   & \textbf{93.61} & -                                                     & 93.85  \\ \bottomrule  
\end{tabular}}
\end{table}

\begin{table}[!ht]
\small
    \centering
    \caption{\small The 4 attributes of the WDC benchmarks used in training \system\ and DM according to \protect\cite{Primpeli:2019:WDC}.}
    \label{tab:wdcattr}
    \begin{tabular}{cp{5.6cm}c} \toprule
    Attributes & Examples & \%Available \\ \midrule
    Title & {Corsair Vengeance Red LED 16GB 2x 8GB DDR4 PC4 21300 2666Mhz dual-channel Kit - CMU16GX4M2A2666C16R Novatech} & 100\% \\ \midrule
    Description & {DDR4 2666MHz C116, 1.2V, XMP 2.0 red-led, Lifetime Warranty
} & 54\% \\ \midrule
    Brand & {AMD} & 19\% \\ \midrule
    SpecTable & Memory Type DDR4 (PC4-21300) Capacity 16GB (2 x 8GB) Tested Speed 2666MHz Tested Latency 16-18-18-35 Tested Voltage 1.20V Registered / Unbuffered Unbuffered Error Checking Non-ECC Memory Features - red-led XMP 2.0 & 7\% \\ \bottomrule
    \end{tabular}
\end{table}

\section{Experiments on Different WDC product attributes} \label{sec:wdc}

Following the settings in \cite{Primpeli:2019:WDC} for the evaluated models, 
we evaluate \system\ on 4 different subsets of the product attributes as input
so that \system\ and DeepMatcher are evaluated under the same setting.
We list the 4 attributes in Table \ref{tab:wdcattr}. Note that except for \textsf{title},
the attributes can be missing the the data entries. For example, the \textsf{SpecTable}
attribute only appears in 7\% of the entries in the full training set.

\begin{table*}[!t]
\caption{F1 scores of \system\ on the WDC datasets with different subsets of the product attributes}
\label{tab:wdcattrexp}
\begin{tabular}{c|cccc|cccc|cccc|cccc} \toprule
          & \multicolumn{4}{c|}{title}       & \multicolumn{4}{c|}{title\_description} & \multicolumn{4}{c|}{title\_description\_brand} & \multicolumn{4}{c}{title\_description\_brand\_specTable} \\
          & small & medium & large & xlarge & small   & medium   & large   & xlarge  & small     & medium     & large    & xlarge    & small        & medium       & large       & xlarge       \\ \midrule
all       & 84.36 & 88.61  & 93.05 & 94.08  & 69.51   & 75.91    & 81.56   & 87.62   & 68.34     & 75.43      & 84.80    & 85.19     & 67.08        & 75.55        & 83.08       & 84.44        \\
cameras   & 80.89 & 88.09  & 91.23 & 93.78  & 61.64   & 73.41    & 79.51   & 83.61   & 59.97     & 73.16      & 78.60    & 82.61     & 55.04        & 68.81        & 76.53       & 80.09        \\
computers & 80.76 & 88.62  & 91.70 & 95.45  & 66.56   & 75.60    & 87.39   & 92.26   & 65.15     & 73.55      & 86.05    & 90.36     & 60.82        & 66.90        & 84.25       & 88.45        \\
shoes     & 75.89 & 82.66  & 88.07 & 90.10  & 59.57   & 69.25    & 76.33   & 76.27   & 57.43     & 71.57      & 77.07    & 77.39     & 56.57        & 71.02        & 76.58       & 75.63        \\
watches   & 85.12 & 91.12  & 95.69 & 96.53  & 58.16   & 70.14    & 81.03   & 84.55   & 59.66     & 73.06      & 81.92    & 84.46     & 52.49        & 68.67        & 79.58       & 82.48    \\ \bottomrule
\end{tabular}
\end{table*}

We summarize the results in Table \ref{tab:wdcattrexp}. Among all the tested combinations
(the same as the ones tested for DeepMatcher in \cite{Primpeli:2019:WDC}), the combination
consisting of only the \textsf{title} attribute works significantly better than the others.
The difference ranges from 3.2\% (computer, xlarge) to over 30\% (watches, small).
According to this result, we only report \system's results on the \textsf{title} attribute while
allowing DeepMatcher to access all the 4 attributes to ensure its best performance.

% 342.7118115005183
The performance of \system\ drops when more attributes are added is because of the
sequence length. For example, for the combination \textsf{title+description}, we found that the 
average sequence length grows from 75.5 (\textsf{title} only) to 342.7 
which is beyond our default max length of 256 tokens.
As a results, some useful information from the \textsf{title} attributes is
removed by the summarization operator.

\end{document}